\documentclass[aps,prb,superscriptaddress,twocolumn]{revtex4-2}

\usepackage{setspace}
\usepackage{lmodern}

\usepackage[latin9]{inputenc}
\usepackage{amsmath}
\usepackage{amssymb}
\usepackage{graphicx}
\usepackage[colorlinks=True,linkcolor=red,citecolor=blue,urlcolor=blue]{hyperref}
\usepackage{epstopdf}
\usepackage{color}
\usepackage{enumerate}
\usepackage{ulem}
\usepackage{bm} 
\usepackage{physics}

\usepackage{array}
\usepackage[labelfont=bf]{caption}
\usepackage{multirow}
\usepackage{floatpag}
\usepackage[misc]{ifsym}
\usepackage{sistyle}
\usepackage{upgreek}
\usepackage{xcolor}

\begin{document}

\title{Chiral-helical junctions in screened graphene}

\author{Bilal Kousar}
\author{Selma Franca}
\author{David Perconte}
\affiliation{Univ. Grenoble Alpes, CNRS, Grenoble INP, Institut N\'{e}el, 38000 Grenoble, France}
\author{Anton Khvalyuk}
\affiliation{LPMMC, Univ. Grenoble Alpes, 38000 Grenoble, France}
\author{Wenmin Yang}
\affiliation{Univ. Grenoble Alpes, CNRS, Grenoble INP, Institut N\'{e}el, 38000 Grenoble, France}
\author{Hadrien Vignaud}
\affiliation{Univ. Grenoble Alpes, CNRS, Grenoble INP, Institut N\'{e}el, 38000 Grenoble, France}
\author{Fr\'{e}d\'{e}ric Gay}
\affiliation{Univ. Grenoble Alpes, CNRS, Grenoble INP, Institut N\'{e}el, 38000 Grenoble, France}
\author{Kenji Watanabe}
\affiliation{Research Center for Electronic and Optical Materials, National Institute for Materials Science, 1-1 Namiki, Tsukuba 305-0044, Japan}
\author{Takashi Taniguchi}
\affiliation{Research Center for Materials Nanoarchitectonics, National Institute for Materials Science,  1-1 Namiki, Tsukuba 305-0044, Japan}
\author{Clemens B. Winkelmann}
\affiliation{Univ. Grenoble Alpes, CEA, Grenoble INP, IRIG-Pheliqs, Grenoble, France}
\author{Yangtao Zhou}
\affiliation{Liaoning Academy of Materials, Shenyang 110167, P. R. China}
\author{Zheng Vitto Han}
\affiliation{State Key Laboratory of Quantum Optics Technologies and Devices, Institute of Optoelectronics, Shanxi University, Taiyuan 030006, P. R. China}
\affiliation{Collaborative Innovation Center of Extreme Optics, Shanxi University, Taiyuan 030006, P. R. China}
\affiliation{Liaoning Academy of Materials, Shenyang 110167, P. R. China}
\author{Alexandre Assouline}
\affiliation{Univ. Grenoble Alpes, CNRS, Grenoble INP, Institut N\'{e}el, 38000 Grenoble, France}
\author{Jens H. Bardarson}
\affiliation{Department of Physics, KTH Royal Institute of Technology, 106 91 Stockholm, Sweden}
\author{Adolfo G. Grushin}
\affiliation{Univ. Grenoble Alpes, CNRS, Grenoble INP, Institut N\'{e}el, 38000 Grenoble, France}
\affiliation{Donostia International Physics Center (DIPC),
Paseo Manuel de Lardiz\'{a}bal 4, 20018, Donostia-San Sebasti\'{a}n, Spain}
\affiliation{IKERBASQUE, Basque Foundation for Science, Maria Diaz de Haro 3, 48013 Bilbao, Spain}
\author{Hermann Sellier}
\author{Benjamin Sac\'{e}p\'{e}}
\affiliation{Univ. Grenoble Alpes, CNRS, Grenoble INP, Institut N\'{e}el, 38000 Grenoble, France}

\begin{abstract}
\textbf{
Reproducibility and quantization in quantum spin Hall platforms is a persisting challenge, limiting their use in hybrid realizations of topological superconductivity. We report robust and reproducible quantized transport in a graphene quantum Hall topological insulator, stabilized at low magnetic fields by screening long-range Coulomb interactions with a metallic Bi$_2$Se$_3$ back gate. Beyond quantized resistance plateaus, we demonstrate mode-resolved control via gate-defined chiral-helical junctions that selectively transmit or backscatter a single helical channel, a capability inaccessible in time-reversal symmetric quantum spin Hall systems. Targeted experiments and simulations identify contact-induced doping, effectively creating unintended chiral-helical interfaces, as a generic mechanism for quantization breakdown, which is mitigated by large area contacts that enhance edge-channel equilibration. Our findings establish metal screened graphene as a gate-tunable, interaction-driven helical system with quantized transport, spatially separable helical channels, and compatibility with superconducting proximity for topological devices.
}
\end{abstract}

\maketitle

\subsection*{Introduction} 

Helical edge states---one-dimensional channels in which opposite spins counter-propagate---provide a natural setting for realizing topological quantum phases. When coupled to superconductors and gapped by appropriate symmetry-breaking fields, such edges are predicted to host Majorana zero modes~\cite{Fu08,Fu09}, a key ingredient for fault-tolerant quantum computation~\cite{Nayak08,Hasan10,Qi2011}. Despite this appealing framework, a platform that reliably supports clean, gate-tunable helical edges has remained elusive~\cite{Yazdani23}.

A viable system must meet two stringent requirements. First, helical edge transport must be demonstrated through reproducible signatures, including quantized conductance and the exclusion of bulk or trivial edge contributions. Second, one must be able to locally open and control a gap in the helical spectrum---ideally via electrostatic gating or magnetic field---to create domain boundaries where Majorana modes can be localized. Existing implementations based on time-reversal-symmetric topological insulators~\cite{Rokhinson2012,Wiedenmann2016} or semiconductor-superconductor hybrids~\cite{Mourik2012} have not satisfied both conditions simultaneously, leaving the field without a robust, experimentally accessible helical-edge platform.

 \begin{figure*}[ht!]
            \centering
 \includegraphics[width=0.9\linewidth]{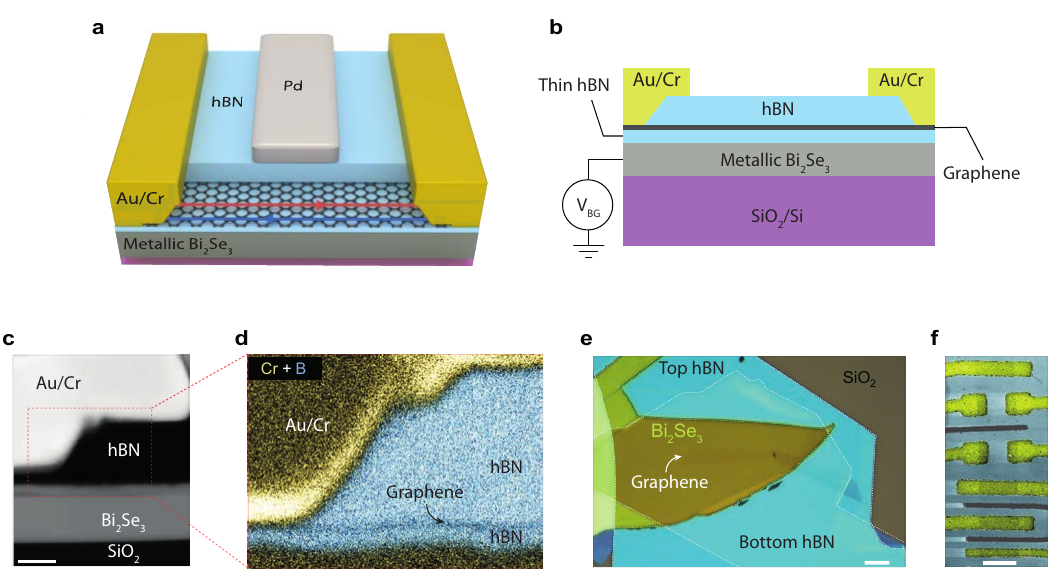}
 \caption{\textbf{Metallic screening architecture for stabilizing the graphene quantum Hall topological insulator.}
    \textbf{a.} Schematic of the device architecture. Monolayer graphene is encapsulated between thin bottom and thick top hBN layers, placed atop a metallic Bi$_2$Se$_3$ flake that serves as a proximal backgate to screen long-range Coulomb interactions. Planar Au/Cr contacts are fabricated by etching the top-hBN to avoid short-circuits through the thin hBN spacer. The device is equipped with a Pd top-gate.
    \textbf{b.} Schematic cross-section of the layered stack of graphene, hBN, and Bi$_2$Se$_3$. 
    \textbf{c.} Transmission electron micrograph of a representative cross-section. Scale bar is 20 nm. 
    \textbf{d.} Magnified electron energy loss spectroscopy (EELS) spatial map confirming the layer structure and planar contact to graphene.
    \textbf{e.} Optical image of a representative heterostructure, showing bubble-free interfaces. Scale bar is 2 $\upmu$m.
    \textbf{f.} False-color atomic force microscopy (AFM) images of the devices, with Au/Cr contacts in yellow and Pd top gates in grey. Scale bar is 2 $\upmu$m. e and f correspond to sample BK82 (see SI Table I).}
 \label{fig1}
  \end{figure*}

The quantum Hall topological insulator (QHTI) phase~\cite{Abanin06,Fertig06,Kharitonov16}, which arises from SU(4) symmetry breaking in the zeroth Landau level of graphene~\cite{Abanin06,Fertig06,Alicea06,Yang06,Herbut07,Kharitonov12,Wei25,An25}, offers a compelling alternative.
By combining the topological protection of the quantum Hall effect and a U(1) spin symmetry resulting from strong interaction-driven spin ordering, QHTIs support a tunable regime of helical edge transport ~\cite{Young14,Maher13,Veyrat20,coissard2022a}. A key advantage of QHTIs is that the resulting quantum spin Hall (QSH) effect is rooted in the intrinsically robust physics of Landau levels and their quantum Hall edge channels, which have recently been shown to support superconducting coherence when proximitized in suitably engineered devices~\cite{Amet16,Lee2017,Zhao2020a,Gul2022,Vignaud23}. 

QHTIs further enable a unique functionality not accessible in time-reversal-symmetric QSH systems: the controlled formation of chiral-helical junctions that interface the QHTI state with the adjacent quantum Hall phase. Using local gates, one can selectively transmit or backscatter a single spin-polarized branch of the helical edge along such a junction, providing direct, mode-resolved control, and enabling interferometry~\cite{wei2017,Jo21} and symmetry-tunable topological Josephson devices~\cite{Blasi19,Blasi23}.

Despite these advantages, existing approaches to induce the QHTI phase face their own specific challenges. 
Tilted magnetic fields can drive the transition to the QHTI phase~\cite{Young14,Maher13}, but require fields exceeding 30 T in monolayer graphene and 15 T in bilayers, incompatible with superconducting device integration. High dielectric constant substrates such as SrTiO$_3$~\cite{Couto11} can screen long-range Coulomb interactions and stabilize the QHTI at lower magnetic fields~\cite{Veyrat20,coissard2022a}, but suffer from non-linear dielectric response~\cite{Veyrat20}, hysteresis, temporal drift, domain formation~\cite{Frenkel17}, and uncontrolled in-plane electric fields~\cite{Csonca25}, limiting device stability and reproducibility.

Here, we engineer a robust and scalable platform by integrating exfoliated metallic Bi$_2$Se$_3$ flakes as a proximal gate beneath monolayer graphene, separated by a sub-10nm thick hexagonal boron nitride (hBN) bottom spacer. Bi$_2$Se$_3$ stands out as one of the few exfoliable metallic layered materials that provides a flat surface, remains non-superconducting down to millikelvin temperatures, and thus does not screen the applied perpendicular magnetic field. This architecture shown in Fig.~\ref{fig1}a and b enables efficient screening of Coulomb interactions, stable and non-hysteretic gating, and compatibility with superconducting devices. 
To prevent short-circuiting through the ultrathin hBN spacer, we precisely etch the top hBN and employ planar contacts.
Using this platform, we demonstrate highly reproducible, quantized helical edge transport at low magnetic fields and provide direct evidence for spin-resolved edge channel control through gate-defined chiral-helical junctions. 

 \begin{figure*}[ht!]
            \centering
 \includegraphics[width=1\linewidth]{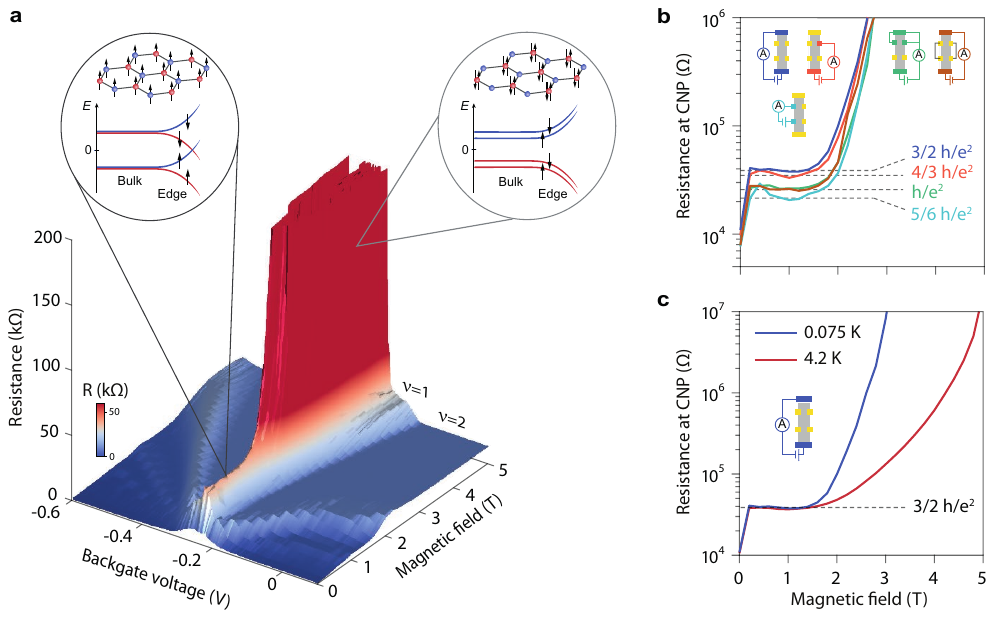}
 \caption{\textbf{Helical edge transport in screened graphene.}
    \textbf{a.} Two-terminal resistance of the Hall bar device in Fig. 1\textbf{f} (sample BK82) with top gate set to iso-density, plotted as a function of magnetic field and back-gate voltage at $T = 0.075$~K, showing quantized resistance plateau at charge neutrality associated with helical edge transport.
    \textbf{b.}  Two-terminal resistance curves at the charge neutrality point (CNP) as a function of magnetic field at 0.075~K for the configurations in the inset.
    \textbf{c.} Two-terminal resistance at CNP measured at  4.2~K and 0.075~K, showing no temperature dependence  of the QHTI plateau, and insulating behavior in the valley polarized phase.}
 \label{fig2}
  \end{figure*} 
  
\subsection*{QHTI phase in Bi$_2$Se$_3$-screened graphene}

Figures~\ref{fig1}e and f show a bubble-free van der Waals heterostructure comprising hBN-encapsulated graphene with a 6.5-nm-thick bottom spacer on Bi$_2$Se$_3$, along with representative devices including a Hall bar and two-terminal devices equipped with Au/Cr contacts (yellow electrodes) and Pd top gates (dark gray electrodes). Cross-sectional transmission electron micrograph and EELS mapping (Fig.~\ref{fig1}c and d) confirm the stacked architecture, including surface contacts to graphene, the hBN spacer between graphene and Bi$_2$Se$_3$ and a 7.4 nm thin native oxide atop Bi$_2$Se$_3$.

We first establish the robustness of the helical phase in this screened architecture (see Fig.~\ref{fig2}). The helical phase manifests itself in Fig.~\ref{fig2}a as a magnetic-field-independent resistance plateau at charge neutrality, separating the filling factor $\nu=\pm 2$ QH plateaus. 
The resistance is quantized to $3/2\,\times h/e^2$, as expected for the Hall bar device in \ref{fig1}f~\cite{konig07}, where the number of helical edge segments connecting contacts on the left and right edges of the device determines the two-terminal resistance~\cite{Young14,Veyrat20}. As shown in Fig.~\ref{fig2}b, varying the contact configuration changes the number of helical segments between the source and the drain, yielding distinct resistance plateaus in excellent agreement with the expected quantized values (dashed black lines). This quantization is reproducible across devices of different sizes and from different samples (see SI).

\begin{figure*}[ht!] 
\includegraphics[width=0.75\linewidth]{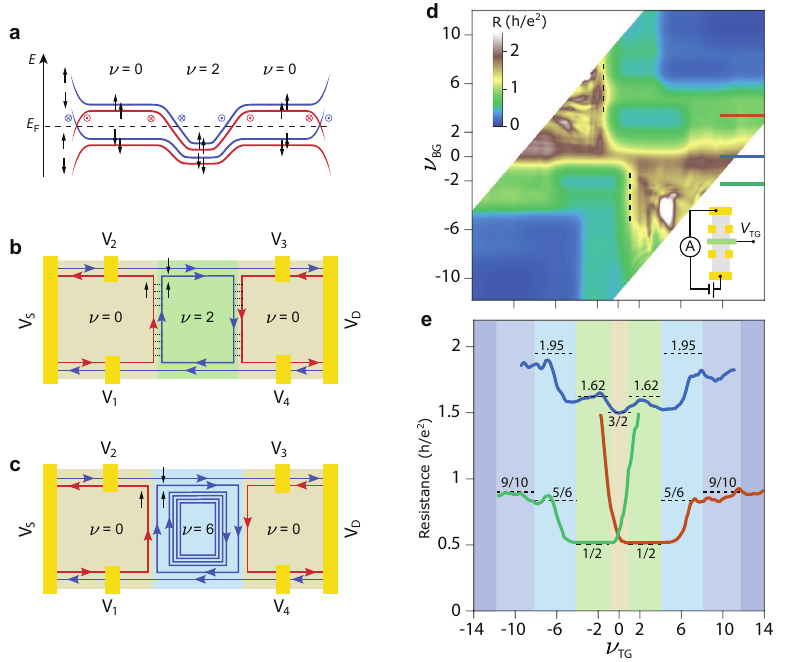}
\centering
\caption{\textbf{Selective backscattering and equilibration in chiral-helical junctions.} \textbf{a.} Landau level dispersion for \(N=0\)  across a junction where the filling factor \(\nu\) changes from 0 in the region with only the back gate to 2 in the region controlled by both back gate and top gate. At \(\nu=0\), we assume a spin-split zeroth Landau level with spin-polarized helical edge states, while the gaps in the  $\nu = 2$ region are negligible (the small lifting of the 4-fold degeneracy is just for visualization purpose).
\textbf{b.} Schematic of a chiral-helical junction in a six-terminal Hall bar with top-gated region set to \(\nu = 2\). Voltage probes \(V_{1-4}\) and source-drain current leads \(V_S\), \(V_D\) are indicated. The backscattered hole-like helical mode equilibrates along the interface via spin-selective intermode scattering indicated by local tunneling bridges (dashed lines), thereby increasing conductance. \textbf{c.} Schematic of a chiral-helical junction with top-gated region set to \(\nu = 6\). In addition to equilibration along the interfaces, spin-selective equilibration between the spin-down electron-like helical edge mode and the \(N=1\) spin-down chiral modes is possible, inducing backscattering and increasing resistance.    
\textbf{d.} Colormap of the two-terminal resistance measured in the configuration shown in the inset at 1.8~T, as a function of the top-gated region filling factor \(\nu_{\mathrm{TG}}\) and the back gate-only region filling factor \(\nu_{\mathrm{BG}}\) for sample BK47 (see SI). The blue, orange, and green lines correspond to fixed back gate filling factors \(\nu_{\mathrm{BG}}=0\), \(\nu_{\mathrm{BG}}=+2\), and \(\nu_{\mathrm{BG}}=-2\), respectively. \textbf{e.} Linecuts of the two-terminal resistance at fixed \(\nu_{\mathrm{BG}}\) in panel d. Dashed lines indicate expected values for different filling factor combinations (see Methods).}
\label{fig3}
\end{figure*} 

A second hallmark of helical edge transport is its metallic character, which rules out the presence of a small gap in the edge excitations.  Fig.~\ref{fig2}c shows linecuts at $\nu=0$ extracted from Fig.~\ref{fig2}a for two different temperatures, 4.2 and 0.075 K. While for $B\gtrsim 1.8$ T, the insulating phase shows an activated increase of resistance upon cooling, the quantized plateau at lower $B$ is clearly temperature-independent, consistent with gapless, topologically protected helical edges. 

In contrast to SrTiO$_3$ substrates~\cite{Veyrat20}, Bi$_2$Se$_3$ yields a perfectly linear field effect, enabling stable and reproducible electrostatic tuning of the QHTI phase. Its metallic character efficiently screens long-range Coulomb interactions in graphene for electron separations exceeding the hBN spacer thickness, that is, when $l_B > d_{\rm BN}$. 

The mechanism underpinning the emergence of the QHTI phase in screened graphene is a unique screening-tunable renormalization of the anisotropy energies that govern spin and valley symmetry breaking in the zeroth Landau level at half filling~\cite{Veyrat20}. These anisotropy energies arise from the Zeeman effect, electron-phonon interaction, lattice-scale short-range Coulomb interactions and Moir\'e potential, which breaks the approximate SU(4) symmetry, and favor spin or valley polarization, or a superposition of both~\cite{Alicea06,Yang06,Herbut07,Kharitonov12,Wei25,An25}. 

The role of the long-range Coulomb interaction is central: first, it drives the formation of a large quantum Hall ferromagnetism gap in the zeroth Landau level, with a magnitude that depends on magnetic field and dielectric environment as evidenced by tunneling spectroscopy~\cite{coissard2022a}; second, it renormalizes the valley anisotropy energies, thereby critically shifting the delicate balance between competing symmetry-broken phases~\cite{Kharitonov12,Wei25}. While a modest change in the long-range Coulomb interaction---via a change in the dielectric environment---can tune from the Kekul\'e bond order to a canted antiferromagnetic order~\cite{Kharitonov12,Wei25}, or even a coexistence of both~\cite{An25}, strong screening by either a high dielectric constant substrate~\cite{Veyrat20} or, as demonstrated here, from the metallic Bi$_2$Se$_3$ back gate, effectively suppresses this renormalization. 
This suppression ultimately shifts the energy landscape in favor of spin polarization, as supported by ab-initio calculations~\cite{Wei25}, stabilizing the topological ferromagnetic phase, that is, the QHTI~\cite{Kharitonov16}, with robust helical edge states as observed in our experiments.

Importantly, the QHTI phase transitions at $B \sim 2$\,T into a valley-polarized insulating phase (see diverging resistance ridge at $B \gtrsim 2$\,T in Fig.~\ref{fig2}a), identified by scanning tunneling microscopy as a sublattice-polarized charge density wave~\cite{coissard2022a}. This transition is driven by the reduction in screening efficiency at higher magnetic fields, where $l_B < d_{\rm BN}$. Although the order of the transition, first or second, remains unknown, the associated symmetry breaking naturally opens a magnetic-field tunable gap in the helical spectrum, a critical ingredient for engineering controllable topological Josephson junctions~\cite{Fu08,SanJose15}.

\subsection*{Selective backscattering of a helical edge state}

A key novelty of the QHTI phase is the ability to directly probe helical edge states by interfacing them with chiral quantum Hall edge channels. 
This additional functionality in our device architecture is made possible by electrostatic top gate electrodes (see Fig.~\ref{fig1}f).  
As illustrated in Fig.~\ref{fig3}a-c, the top gate enables local modulation of the filling factor beneath it, while maintaining charge neutrality in the outer regions in the QHTI phase, thus forming 0$n$0 or 0$p$0 junctions. By tuning the central region to filling factors $\pm 2$, we realize a chiral-helical junction that selectively backscatters one of the two counterpropagating helical edge channels while transmitting the other (see Fig.~\ref{fig3}a,b).This capability provides a distinct and previously inaccessible probe of helical edge states in QSH systems.

Figure~\ref{fig3}d presents a two-dimensional resistance map as a function of top gate and back gate filling factors, $\nu_{\mathrm{TG}}$ and $\nu_{\mathrm{BG}}$, respectively, for the configuration shown in the inset of Fig.~\ref{fig3}d. The resistance map displays characteristic rectangular plateaus, typical for chiral quantum Hall $npn$ junctions~\cite{Williams07,Amet13} (see the red and green line cuts in Fig.~\ref{fig3}e, demonstrating typical equilibration between chiral quantum Hall channels). However, unlike unscreened graphene devices~\cite{Amet13}, the two bipolar quadrants ($\nu_{\mathrm{BG}}.\nu_{\mathrm{TG}}<0$) do not align at $\nu_{\mathrm{TG}}=0$. Instead, their horizontal offset, highlighted by two black dashed lines, reveals the presence of a QHTI gap at $\nu_{\mathrm{TG}}=0$ under the top gate. For comparison, an insulating graphene device shows a positive gap at $\nu=0$ with well-separated quadrants (see SI Fig.~\ref{figS3}).

When the bulk is tuned to the QHTI phase ($\nu_{\mathrm{BG}}=0$), adjusting the top gate away from neutrality enables selective backscattering of helical modes. As shown by the blue linecut $\nu_{\mathrm{BG}}=0$ in Fig.~\ref{fig3}e extracted from Fig.~\ref{fig3}d, for $\nu_{\mathrm{TG}}=0$ the resistance reaches the expected helical quantized value of $3/2\times h/e^2$. Shifting to $\nu_{\mathrm{TG}}=\pm 2$ induces backscattering of one helical edge mode, while the other is fully transmitted, resulting in a resistance plateau close to the expected value of 1.62 $h/e^2$. 

To quantify these observations, we consider the equilibration between the backscattered helical channel (red channel in Fig.~\ref{fig3}b) and the co-propagating chiral inner channel beneath the top gate (blue channel), which share the same spin polarization (see Landau level dispersion in Fig.~\ref{fig3}a). Assuming a fictitious voltage probe within Landauer-B\"uttiker formalism to model this spin-selective equilibration~\cite{Amet13}, we calculate theoretical resistance values (see Methods), indicated by black dashed lines in Fig.~\ref{fig3}e. The experimental values show excellent agreement with these predictions at $\nu_{\mathrm{TG}}=\pm 2$ and remain consistent, albeit slightly reduced, at $\nu_{\mathrm{TG}}=\pm6$ , most likely due to partial equilibration between edge states of the $N=0$ and $N=1$ Landau levels (see Fig.~\ref{fig3}c). 

This controlled backscattering of individual helical edge states, quantitatively captured by a spin-selective equilibration model, provides direct and definitive experimental evidence of helical edge states in the QHTI phase of screened graphene.

\subsection*{Failed resistance quantization due to contact doping}

\begin{figure*}[tb!]
\includegraphics[width=1\linewidth]{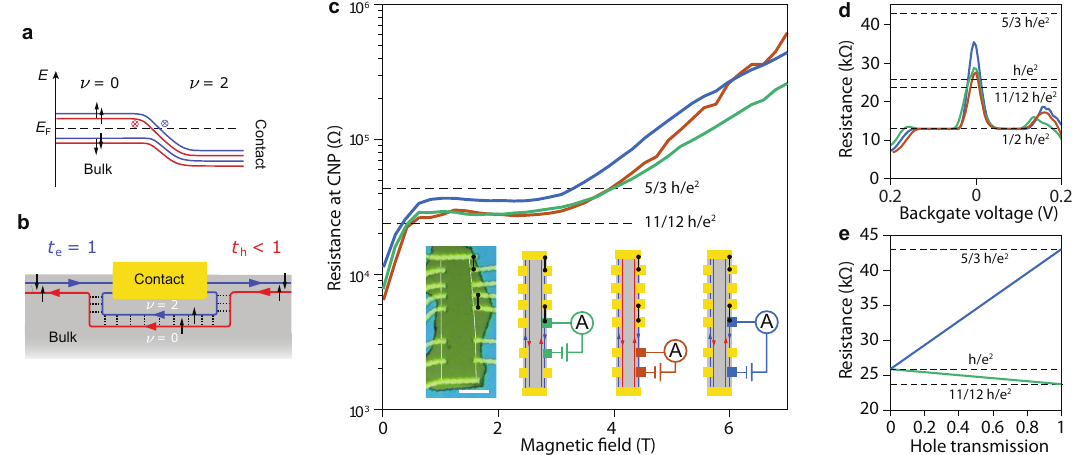}
\caption{\textbf{Asymmetric contact transmission due to induced doping} \textbf{a.} Deviation of the filling factor from \(\nu = 0\) between the bulk and the contact vicinity, caused by n-type doping from the contact. 
\textbf{b.} Schematic illustrating the contact doping effect, which creates additional electron channel near the contact and prevents the hole-like helical edge mode from entering directly. Partial entry can occur via spin-selective inter-mode scattering (dashed lines), resulting in a transmission \(t_{\rm h} < 1\). Electrons fully enter the contacts with transmission \(t_{\rm e} = 1\).
\textbf{c.} Two-terminal resistance at charge neutrality as a function of magnetic field, measured on the multi-terminal device BK41 shown in the inset (scale bar: 5~\(\mu\)m) with a bottom hBN spacer of 3.3 nm. The contact configurations are color-coded and illustrated in the inset schematics (black line connections between contacts indicate contacts shorted during fabrication). Dashed lines mark the expected resistance values for helical edge transport: $11/12 \times h/e^2$ for nearest-neighbor contacts (green and red curves) and $5/3 \times h/e^2$ for second-nearest neighbors (blue curve). \textbf{d.} Two-terminal resistance linecuts as a function of back-gate voltage for the configurations in panel~\textbf{c}, measured at a fixed magnetic field of 2.4~T. \textbf{e.} Two-terminal helical edge resistance for nearest-neighbor (green) and second-nearest-neighbor (blue) contacts, calculated using the Landauer-B\"uttiker formalism (see Methods) as a function of hole transmission \(t_{\rm h}\).}
\label{fig4}
\end{figure*} 

A persistent challenge in QSH systems is the poor quantization of resistance~\cite{konig07,Roth09,Maher13,Fei17,wu18,Kang2024,Ghiasi2025}, which has long cast doubt on the very existence of helical edge states. Various scenarios were put forth to account for the deviation from quantization, including bulk transport or backscattering between helical edge states via nearby spinful bulk impurities or puddles~\cite{Maciejko09,Schmidt12,Vayrynen13}. Here we identify a ubiquitous and intrinsic problem: the incompatibility between the charge neutral nature of QSH states and the unavoidable charge transfer by metallic contact electrodes, which leads to asymmetric transmission at the contacts between electron-like and hole-like helical edge states.

In our devices, charge transfer from the contacts into the graphene induces a $n$-doped region around each contact (Fig.~\ref{fig4}a). This forms a local chiral-helical junction that diverts the hole-like channel from the contact, as illustrated in Fig.~\ref{fig4}b. 
Consequently, while the electron-like channel can be fully transmitted ($t_{\rm e}\simeq 1$), the hole-like channel has a reduced transmission ($t_{\rm h}<1$), determined by disorder and by local equilibration between quantum Hall channels (indicated by dotted lines in Fig.~\ref{fig4}b).

The consequence of this asymmetric edge channel absorption by the contact is illustrated in Fig.~\ref{fig4}c, which shows two-terminal resistance at charge neutrality as a function of magnetic field for a device with multiple small contacts (see inset). To highlight this effect, we compare the resistance measured between two adjacent helical segments (green and red configurations in Fig.~\ref{fig4}c and d) that share a common contact, and the resistance of both sections measured in series (blue configuration). Interestingly, despite differing in length by a factor of two, the individual segments exhibit nearly identical resistance plateaus of value 26.8~k$\Omega$---consistent with helical edge transport but with values higher than the expected QSH quantization of $11/12 \times h/e^2 \simeq 23.6$~k$\Omega$. On the other hand, the total two-segment resistance reveals a value of 36~k$\Omega$
, which is well below the expected QSH value of $5/3 \times h/e^2 \simeq 43~k\Omega$.

This counterintuitive behavior can be understood qualitatively within a simple Landauer-B\"uttiker model that assumes identical asymmetric transmission $t_{\rm e}=1$ and  $t_{\rm h} < 1$ for all contacts (see Methods). Fig.~\ref{fig4}e shows the resulting resistance for helical edge transport as a function of $t_{\rm h}$. We see that the resistance of individual helical segments is quantized at $h/e^2$ for  $t_{\rm h} =0$ (see green curve), indicating transport through a single chiral (electron-type) channel, and then drops towards the expected QSH value reached at $t_{\rm h} = 1$. Similarly, the total two-section resistance also begins at $h/e^2$ and increases linearly with $t_{\rm h}$ to reach the QSH value. This simplified model is consistent with our data: it yields a single-section resistance higher than the expected quantized value, and a two-section resistance lower than the expected value. 

In light of the contact-related transmission asymmetry, it is worth revisiting the data in Fig.~\ref{fig2} and Fig.~\ref{fig3}, which, in contrast, exhibit ideal QSH resistance quantization. The key difference from the sample in Fig.~\ref{fig4} lies in the width of the contacts, which were 3--4 times larger. These larger contacts enable more efficient disorder-induced equilibration between the hole-type edge channel and the $n$-doped region near the contacts, leading to a hole transmission much closer to that of the electron channel. Consequently, the resulting helical edge transport with such large contacts aligns remarkably well with the expected QSH quantization.

\begin{figure}[tb!]
\includegraphics[width=\columnwidth]{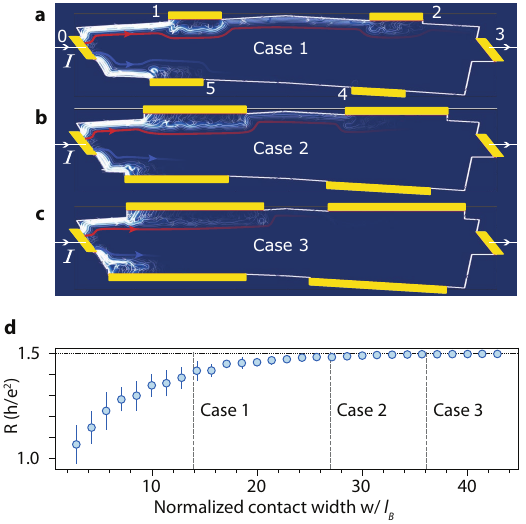}
\caption{
\textbf{Numerical simulations of edge equilibration along the contacts.}
\textbf{a-c} Six-terminal geometry with the scattering region of irregular shape and the electrical contacts represented by golden parallelograms. The scattering region is simulated with a tight-binding model of a graphene flake with Anderson disorder potential, chosen stronger at the contacts than in the bulk to simulate contact induced disorder (see Methods). \textbf{a-c} show the current density for a two terminal resistance experiment between the left- and right-most current leads (labeled 0 and 3, respectively) for three different voltage contact widths $w$.
\textbf{d.} Disorder averaged two-terminal resistance $R$ between leads 0 and 3 as a function of the width of the voltage leads $w$, normalized by magnetic length $l_{B}$. The contact widths in \textbf{a-c} are indicated as vertical dashed lines in \textbf{d}. $R$ reaches the QSH value $3/2 \times h/e^2$ only when the contact width exceeds several tens of $l_B$.}
\label{fig5}
\end{figure}

To further support this hypothesis, we simulate the influence of contact width $w$ on the two-terminal resistance $R_{2t}$ in a disordered QHTI device. To eliminate geometric artifacts, we use an irregularly shaped scattering region, see Fig.~\ref{fig5}{a-c}, including a stronger Anderson potential disorder around the contacts compared to the bulk, and contact-induced gate potential, see Methods. Figs.~\ref{fig5}a-c illustrate the current density for different $w/l_B$, confirming improved absorption of the edge modes into voltage contacts $1$ with increasing $w$. As a consequence, the disorder-averaged two-terminal resistance $R_{2t}$ as a function of $w$ reveals a monotonic rise towards $3/2\times h/e^2$ with increasing contact width, see Fig.~\ref{fig5}d. 
This trend reflects the enhanced equilibration between co-propagating edge channels along the wider n-doped contact region, where the disorder-induced mean free path becomes shorter than $w$~\cite{Buttiker1988,Wen1994,Kane1995}. In the Supplementary Information, we have similarly simulated the multi-terminal geometry in Fig.~\ref{fig4}c, supporting the conclusions drawn from Fig.~\ref{fig4}e.

Our simulations capture the experimental observation that wider contacts lead to improved resistance quantization. 
For instance, using $1.4 \mu$m wide contacts, the correct quantized resistance $3/2 \times h/e^2$ is obtained in Fig. \ref{fig2} at $B = 1$ T ($l_B = 25$ nm) with a normalized contact width $w / l_B = 56$. According to the simulations in Fig. \ref{fig5}d, this value is sufficiently large to enable complete equilibration of the hole-type helical edge channel at the contacts.

These results demonstrate that contact width plays a critical role in enabling proper spin mixing and chemical potential equilibration, thereby restoring the quantized resistance expected for QSH transport.

 \section*{Conclusion and Outlook}

Our work establishes QHTI in metal-screened graphene as a robust platform for realizing helical edge states. The ability to tune the ground state of the half-filled zeroth Landau level with magnetic field, from a ferromagnetic phase supporting helical edge modes to a charge-density-wave  insulating phase, opens the path toward controllably gaping edge excitation. Such \textit{in-situ} symmetry-breaking capability is essential for localizing topological excitations in helical Josephson junctions~\cite{Fu08,SanJose15}. 
In addition, gate-defined chiral-helical junctions provide a new functionality for spin-selective edge transmission and conclusive evidence for the existence of helical edge channels.

At the same time, the contact-induced transmission asymmetry we identify introduces a critical challenge. It not only necessitates a careful re-evaluation of prior helical edge transport measurements, but also poses a significant obstacle for the realization of helical Josephson junctions and other superconducting devices. Specifically, the doped region near metallic contacts breaks the charge neutrality of QSH systems, resulting in unequal transmission between electron-like and hole-like edge channels. In QSH-based superconducting devices, this asymmetry can strongly suppress local Andreev reflections for helical edge states -- a problem to be addressed theoretically.

As in semiconducting platforms with large spin-orbit coupling, a central dilemma emerges: the need to simultaneously achieve efficient superconducting proximity while preserving low doping or charge neutrality in the topological system. Addressing this tension will require innovative device strategies --such as spatially selective gating, engineered tunnel barriers, or novel contact materials-- that preserve helical symmetry at contacts and enable robust proximity effects.

\textit{Note: A very recent work \cite{Domaretskiy25} that appeared during the preparation of our manuscript confirmed the existence of the QHTI phase in graphene screened with a graphite gate.}

\section*{Methods}

\subsection*{Sample fabrication}
Graphene and hBN flakes were mechanically exfoliated onto oxidized silicon wafers, and suitable flakes (typically 30 nm for the top hBN and 5--7 nm for the bottom hBN) were identified by optical and atomic force microscopy. hBN/graphene/hBN heterostructures were assembled using a polycarbonate stamp using the dry transfer technique~\cite{Wang13}. A crucial step during stacking is the intentional misalignment of the hBN flakes with respect to the pristine graphene edges. The hBN crystal axes were misaligned from those of graphene by more than ~15$^\circ$ to avoid moir\'{e} superlattice effects, which can open a gap at charge neutrality and drive the system into an insulating ground state (see Supplementary Information). The resulting heterostructure was deposited onto a Bi$_2$Se$_3$ flake, which serves as both a back gate and a screening electrode. Finally, polymer residues from the transfer process were removed using either oxygen plasma or an AFM tip operated in contact mode, yielding atomically smooth surfaces crucial for controlled etching of the top hBN and for achieving high graphene mobility.

To define contacts, the top hBN was selectively etched by reactive ion etching (RIE) using a CHF$_3$/O$_2$ gas mixture to expose the graphene edges for electrical contact. Prior to device processing, dedicated calibration tests were performed on reference hBN flakes to precisely determine the etch rate, allowing control down to $\sim$1 nm. This ensures etching of only the top hBN layer while preserving the thin bottom hBN spacer that insulates the graphene from the metallic Bi$_2$Se$_3$ back gate. This control is essential to prevent unintentional electrical shorting to the back gate. The contact consists of Au/Cr bilayer, while the topgates are fabricated from Pd. The topgate width was chosen to be 300 nm to maintain sufficient distance between contacts for coherent helical edge transport.

\subsection*{Two-terminal resistance of a multi-terminal device in the helical phase}

In the helical state, each section between two contacts has a resistance of $h/e^2$ (assuming perfect absorption by the contacts) \cite{Abanin07b,Young14,Veyrat20}. As a result, the two-terminal resistance of a measurement configuration with $N_L$ and $N_R$ sections along the left (L) and right (R) edges of the device is given by:
\begin{equation}
R_{2t}(N_L,N_R)=\frac{N_L N_R}{N_L+N_R} \frac{h}{e^2}
\label{Eq1}
\end{equation}
For the various configurations shown in Fig.~\ref{fig2}b, we thus have:
$R_{2t}(3,3) = 3/2 \times h/e^2$,
$R_{2t}(2,4) = 4/3 \times h/e^2$,
$R_{2t}(2,2) = h/e^2$ and 
$R_{2t}(1,5) = 5/6 \times h/e^2$.

\subsection*{Calculations for spin-selective transmission and equilibration of helical edge states} 

We first consider the case where the filling factor beneath the topgated region is tuned to $\nu = 2$, while the rest of the sample is at $\nu = 0$. The four-fold degenerate zLL has a small Zeeman spin splitting at $\nu=2$ and a large interaction-induced spin gap at $\nu=0$. The resulting Landau level dispersion in the two regions is shown in Fig.~\ref{fig3}a. Near the Fermi level at the interface, the lifting of valley degeneracy of the spin-up level gives rise to two counter-circulating but co-propagating electron-like (blue) and hole-like (red) interface states with same spin and opposite valleys. The resulting edge and interface modes are drawn in Fig.~\ref{fig3}b. The current leads are labelled $V_S$ and $V_D$ and the voltage probes as $V_1, V_2, V_3, V_4$.

The backscattering of hole-like helical edge mode should result in a higher resistance than expected for uninterrupted helical edge transport. However, this mode can still contribute to conductance by equilibrating/mixing with the co-propagating electron-like mode along the pn interface via inter-mode scattering, as highlighted by the local tunneling bridges shown as dashed lines in Fig.~\ref{fig3}b. In high-mobility samples, where spin and valley degeneracies are lifted due to minimal Landau level broadening, the edge states can be spin- and/or valley-polarized. It is not immediately clear whether modes with opposite spin or valley polarization can mix. This question was experimentally addressed by Amet et al. \cite{Amet13}, who demonstrated that in high mobility samples, modes with opposite spins do not mix, while those with the same spin mix regardless of valley polarization. The robustness of spin polarization can be attributed to the weak spin-orbit coupling in graphene, whereas weak disorder may still lead to intervalley scattering. 

We adopt a similar assumption in our model, proposing that only the modes with the same spin equilibrate, regardless of the valley polarization. The two co-propagating modes along each interface in Fig.~\ref{fig3}b have the same spin but opposite valley polarization and therefore we assume these two modes fully equilibrate by charge and momentum transfer along the interfaces. After equilibration, the two modes reach the same chemical potential along each interface, which we represent by imaginary voltage probes \(V_L^{eq}\) and \(V_R^{eq}\) at the left and right interfaces, respectively.

\vspace{+5pt}
To calculate the two-terminal resistance, it is necessary to determine the voltage at each probe in Fig.~\ref{fig3}b, including the imaginary voltage probes \(V_L^{eq}\) and \(V_R^{eq}\). This can be conveniently done using the Landauer-B\"uttiker formalism \cite{Buttiker1986}, where the voltage at each probe \(V_i\) is expressed as follows:
    \begin{equation}
  V_i=\frac{\sum_{j} T_{j i} V_j }{\sum_{j} T_{j i}}  
  \label{Eq2}
\end{equation}
where $T_{j i}$ is the transmission probability from the j-th electrode to i-th electrode. In the Landauer-B\"uttiker formalism, the edge channels are assumed to be ballistic modes with a resistance of $h/e^2$.  We assume that both electron-like and hole-like modes are fully transmitted into the contacts such that $T_{ij}=1$ for electron-like modes and $T_{ji}=1$ for hole-like modes, for each pair of contacts $(i,j)$ connected by chiral or helical edge modes. Additionally, we set the source voltage to \(V_S = 1\) and the drain voltage to \(V_D = 0\) in Fig.~\ref{fig3}b. The voltage at each probe is then given by:
\begin{equation}
\begin{split}
 V_1 = \frac{1+V_4}{2} \quad,\quad V_2 = \frac{1+V_L^{eq}}{2} \quad,\quad V_3 = \frac{V_2}{2}   \\ 
 V_4 = \frac{V_R^{eq}}{2} \quad,\quad V_L^{eq} = \frac{V_1+V_R^{eq}}{2} \quad,\quad V_R^{eq} = \frac{V_3+V_L^{eq}}{2}  
 \end{split}
 \label{Eq3}
\end{equation}
We solve these equations numerically, which gives the following results:
\begin{equation}
\begin{split}
 V_1 = 0.6154 \quad,\quad V_2 = 0.7692 \quad,\quad V_3 = 0.3846   \\ 
 V_4 =  0.2307 \quad,\quad V_L^{eq} = 0.5384 \quad,\quad V_R^{eq} = 0.4615  
 \end{split}
 \label{Eq4}
\end{equation}
The total current emanating from the source can be found using the current-voltage relationship in the Landauer-B\"uttiker formalism:
\begin{equation}
I_i=\frac{e^2}{h} \sum_j\left(T_{i j} V_i-T_{j i} V_j\right)
\label{Eq5}
\end{equation}
which yields the current $I_s$ flowing out of the source:
\begin{equation}
\begin{split}
I_S = \frac{e^2}{h} \Bigg[(V_S - V_1) + (V_S - V_2)\Bigg]
= 0.6154 \, \frac{e^2}{h} \,  
\end{split}
\label{Eq6}
\end{equation}
Since $V_{SD} = 1$, we obtain the 2-terminal resistance:
\begin{equation}
{
 R_{2t} (\nu_{BG} = 0, \nu_{TG} = \pm 2) = \frac{V_{SD}}{I_S}  = 1.625 \, \frac{h}{e^2} \, }  
 \label{Eq7}
\end{equation}
This value obtained for $\nu_{BG} = 0$ and $\nu_{TG} = 2$ is also valid for $\nu_{TG} = -2$.

We now consider the case where $\nu_{BG} = 0$ and $\nu_{TG} = 6$, as shown in Fig.~\ref{fig3}c. At high magnetic fields, the large cyclotron gap between the zeroth Landau level (zLL) and $N = 1$ Landau level (LL) suppresses equilibration between their edge channels along the pn interface \cite{Amet13}, since the channels are spatially well separated. In our case, however, we operate at lower magnetic fields where the cyclotron gap is still small, making equilibration more likely. Therefore, we assume that along the two interfaces, the hole-like spin-up mode can equilibrate with three electron-like spin-up modes, one originating from the zLL and two from the $N = 1$ LL. In addition, along the edges in the top-gated region, the spin-down mode from the zLL can equilibrate with the two spin-down modes from the $N = 1$ LL. This equilibration along the edges leads to enhanced backscattering and an increase in resistance, in agreement with the experimental results. 

To calculate the expected values, we again model equilibration using imaginary voltage probes at potentials $V_T^{\mathrm{eq}}$ and $V_B^{\mathrm{eq}}$, along the top and bottom edges of the top-gated region, respectively. Using Eq.~\ref{Eq2}, the linear set of equations describing the voltages are represented by the following matrix form:
\begin{equation}
\begin{bmatrix}
1 & 0 & 0 & 0 & 0 & 0 & 0 & -\tfrac{1}{2} \\ 
0 & 1 & 0 & 0 & -\tfrac{1}{2} & 0 & 0 & 0 \\ 
0 & 0 & 1 & 0 & 0 & -\tfrac{1}{2} & 0 & 0 \\ 
0 & 0 & 0 & 1 & 0 & 0 & -\tfrac{1}{2} & 0 \\ 
-\tfrac{1}{4} & 0 & 0 & 0 & 1 & 0 & -\tfrac{3}{4} & 0 \\ 
0 & -\tfrac{1}{3} & 0 & 0 & 0 & 1 & 0 & -\tfrac{2}{3} \\ 
0 & 0 & -\tfrac{1}{4} & 0 & -\tfrac{3}{4} & 0 & 1 & 0 \\ 
0 & 0 & 0 & -\tfrac{1}{3} & 0 & -\tfrac{2}{3} & 0 & 1
\end{bmatrix}
\begin{bmatrix}
V_1 \\ V_2 \\ V_3 \\ V_4 \\ V_{L}^{eq} \\ V_{T}^{eq} \\ V_{R}^{eq} \\ V_{B}^{eq}
\end{bmatrix}
=
\begin{bmatrix}
\tfrac{1}{2} \\ \tfrac{1}{2} \\ 0 \\ 0 \\ 0 \\ 0 \\ 0 \\ 0
\end{bmatrix}.
\label{Eq8}
\end{equation}
Solving this matrix equation for the voltages and subsequently calculating the two terminal resistance yields $R_{2t}(\nu_{BG} = 0, \nu_{TG}=\pm 6)=1.95~h/e^2 \,$.

\subsection*{Improper resistance quantization due to poor hole equilibration in
the contacts: a toy model}

In this section, we propose a simplified model that reproduces imperfect
resistance quantization in our samples as a result of poor equilibration
of the hole-like edge mode with the contacts. Due to the $n$-doping
induced by the contact metal, an additional electron-like channel
forms at the interface (see Fig.~\ref{fig4}b), creating a chiral-helical junction
that prevents the hole-like mode from entering the contact. However,
disorder in the vicinity of the contact induces inter-mode tunneling
(denoted by the dashed lines in Fig.~\ref{fig4}b). Because the
electron- and hole-like modes are now co-propagating, this leads~\citep{Buttiker1988,Kane1995}
to partial equilibration between them, arranging for partial transmission
of the hole-like wave packets into the contact.

As a result, a hole-like wave packet incident on a contact can either
be transmitted into the contact or reflected to the outgoing hole channel.
Moreover, in our model, the electron- and hole-like wave packets are
effectively decoupled outside the contact area, which allows one to describe the total transmission probability matrix $T_{i\leftarrow j}$ between contacts $i$ and $j$ by a sum of the contributions of individual channels:
\begin{equation}
T_{i\leftarrow j}=T_{i\leftarrow j}^{e}+T_{i\leftarrow j}^{h},\label{eq:T-matrix_sum-over-channels-1}
\end{equation}
where the electron transmission matrix $T_{i\leftarrow j}^{e}$ is given by the standard equation for a chiral edge channel, and $T_{i\leftarrow j}^{h}$ is computed below. Within the Landauer-B\"uttiker formalism, the charge conservation for
contact~$i$ then reads
\begin{equation}
\sum_{j}T_{i\leftarrow j}V_{j}-V_{i}\,\sum_{j}T_{i\leftarrow j}+I_{i}R_{Q}=0,\label{eq:LB-charge-conservation-1}
\end{equation}
where $R_{Q}=h/e^{2}$ is the resistance quantum, $V_{i}$ is the
contact chemical potential, and $I_{i}$ is the external current flowing
into the contact from the current source.  
Note that Eq.~(\ref{eq:LB-charge-conservation-1}) is not sensitive to a simultaneous shift of all~$V_{i}$. 

We can predict any two-contact resistance by solving Eq.~(\ref{eq:LB-charge-conservation-1}) with the appropriate choice of $V_{i}$, $I_{i}$. 
For the two-contact resistance $R_{ab}$, we assume that the current and voltage probes are applied
at contacts $a,b$, with $R_{ab}=\left(V_{a}-V_{b}\right)/I_{a}$.
To remove the remaining freedom in the chemical potential, we set
$V_{a}=1,\,\,\,V_{b}=0$, whereas $I_{a},\,I_{b}$ should be found
from Eq.~(\ref{eq:LB-charge-conservation-1}). By construction, the solution satisfies charge conservation and thus has $I_{a}=-I_{b}$.
The remaining contacts $i\neq a,b$ are subject to floating boundary
conditions: $I_{i}=0$, and $V_{i}$ are to be determined.

We then consider the simplest case in which every contact in a system with $N$ contacts is characterized
by identical hole transmission~$t$. The hole transmission matrix~$T_{i\leftarrow j}^{h}$
can be expressed as a direct sum over all possible paths from $j$
to $i$. To this end, let us enumerate
the contacts in the order of the hole propagation. Without loss of generality, consider
a wave packet originating from contact $j=1$. The probability of absorption by a different contact $i$, $N\ge i>1$, is given by
\begin{align}
T_{i\leftarrow1}^{h} & =t\times\left(r^{i-1}+r^{i-1}\times r^{N}+r^{i-1}\times r^{2N}+...\right)\times t\nonumber \\
 & =\frac{t^{2}r^{i-1}}{1-r^{N}},\label{eq:T-i-1_equal-t-1}
\end{align}
where $r=1-t$. The two outer $t$ factors in this expression correspond, respectively, to the probability of the first transmission from contact $j=1$
to the adjacent hole channel, and to the last transmission from the hole
channel into contact~$i$. 
The inner sum in Eq.~(\ref{eq:T-i-1_equal-t-1}) takes into account all paths from contact~$j$ to contact~$i$, and these paths are enumerated by the required number of full revolutions around the system. 
Along a given path, the wave packet must not be absorbed by any intermediate contact, and each avoided absorption adds a factor~$r$ to the contribution of a given path. Similarly, the matrix element for $i=j=1$ is given by
\begin{equation}
T_{1\leftarrow1}^{h}=r+t\times\left(r^{N}+r^{2N}+...\right)\times t=r+\frac{t^{2}r^{N}}{1-r^{N}},
\end{equation}
where the additional first term corresponds to the trivial contribution
of being reflected back into the contact of origin. The sum over
paths now starts from the full revolution, since, once injected
into the system, the wave packet has to revolve at least once before
it has a chance to tunnel back into the same contact. Since all contacts are identical, the matrix elements for $j \neq 1$  can be restored from $T_{i \leftarrow 1}$ by cyclic shift of both indices: $T_{i\leftarrow j}=T_{i-j+1\mod N\,\leftarrow1}$. 

The particular case of $t=1,\,r=0$ restores the standard form of
the $T$ matrix, when a wave packet originating from contact $i$
is fully absorbed by the nearest downstream contact with probability
one:
\begin{equation}
T_{i\leftarrow j}^{h}(t=1)=\begin{cases}
1, & i\text{ is next downstream of }j,\\
0, & \text{otherwise}.
\end{cases}\label{eq:electron-transmission-matrix-1}
\end{equation}
Note that for the chosen contact enumeration, the electron transmission
matrix~$T^{e}=\left[T^{h}(t=1)\right]^{T}$ is obtained by transposing
the one for holes, since the two modes are counter-propagating.

By solving the linear system~(\ref{eq:T-matrix_sum-over-channels-1})-(\ref{eq:electron-transmission-matrix-1})
with respect to $I_{i},\,V_{i}$ we find that, remarkably, all two-contact
resistances are linear in hole transmission~$t$. Therefore, the general result represents a linear interpolation between the two edge cases: Eq.~\eqref{Eq1} for $t=1$
and $R_{ab}=R_{Q}$ (a single chiral mode) for $t=0$:
\begin{equation}
R_{ab}/R_{Q}=1+\left(\frac{N_{L}N_{R}}{N_{L}+N_{R}}-1\right)t.\label{eq:two-contact-resistance_constan-t-value-1}
\end{equation}
where $N_{L},\,N_{R}$ is the total number of edge sections between contacts $a$
and $b$ in the corresponding direction. Eq.~\eqref{eq:two-contact-resistance_constan-t-value-1}
results in the straight lines plotted in Fig.~\ref{fig4}e. While
the linear trend does not hold in a more realistic model reflecting disorder in the values of $t$ of each contact, Eq.~\eqref{eq:two-contact-resistance_constan-t-value-1} qualitatively reproduces the increasing departure from the ideal values given by Eq.~\eqref{Eq1} as the contact quality decreases, here modeled by a decreasing $t<1$.

\subsection*{Simulation details}

To simulate the effect of narrow contacts on the quantization of the two-terminal resistance in the helical phase, we use the \textsc{Python} package \textsc{Kwant}~\cite{Groth2014}. 
First, we define a graphene system of an irregular shape from a starting rectangle of size $L_x \times L_y= 1800a \times 300a$, see Fig.~\ref{fig5}a.  
Here, $a = 1.42~\textup{\AA}$ denotes the lattice constant, and we assume the strength of the interaction-induced Zeeman splitting to be $V_z = 0.1~\rm eV$, a nearest neighbor hopping strength $ t= 2.8~\rm eV$, and a magnetic flux per unit cell $\phi \approx 0.0012 \phi_0 $ with $\phi_0 = h/e$ being the magnetic-flux quantum. 
We assume that the total chemical potential of the scattering region is position dependent, and at site $n$ equals
\begin{equation}\label{eq:onsite}
 \mu^{\rm tot} (x_n,y_n)  =  \mu_0 + \nu^b  (x_n,y_n)+ 
  \eta (x_n,y_n) \big(V_g+  \nu^c  (x_n,y_n) \big),
\end{equation}
The first two terms describe, respectively, the reference chemical potential, $\mu_0 = 0$, and a bulk disorder potential $\nu^b  (x_n,y_n)$. The latter originates from imperfections in the substrate and it is drawn independently for each site from a uniform distribution $[-W^b/2, W^b/2]$.
The last term in Eq.~\eqref{eq:onsite} models the doping effect of the contacts through a smooth window function $\eta (x_n,y_n)$ that tends to zero (one) in the bulk (close to the contacts).
The parameter $V_g = 0.2~\rm eV$ represents the strength of the the gating effect from the contact 
inside the sample, 
and $\nu^c (x_n,y_n)$ is the disorder potential induced by the contacts that is drawn independently for each site from a uniform distribution $[-W^c/2, W^c/2]$. 
In our calculations we choose $W^c = 2.5t$ and $ W^b\approx 0.18t$, i.e, $W^c > W^b$, because we expect the region close to the contacts to be dirtier than the bulk.
More details can be found in the Supplemental Material.  

Fig.~\ref{fig5} illustrates a six-terminal geometry where each metallic contact modeled as a doped disorder-free graphene sheet with $\mu_0^{\rm lead} = 0.9t$ and $\phi = 0$. 
These contacts are enumerated and represented as golden parallelograms in Fig.~\ref{fig5}{a-c}. Here we assume contacts to be of infinite depth. For a finite-size floating contact, a different equilibration mechanism emerges related to the loss of coherence within the contact, as discussed in Ref.~\cite{Roth09}.
In the Supplemental Material we include a figure that shows the real space distribution of $\mu^{\rm tot} (x_n,y_n)$ for one disorder realization where it can be seen how disorder increases in proximity of voltage contacts $i = 1,2,4,5$ and additional simulations relating to the 6- and 12-terminal devices presented in the main text.

\section*{Acknowledgments}

We thank P. Corbae, S. Csonka, L. Glazmann, F. Hassler, A. Singh, C. Stampfer for valuable discussions. 
B.S. has received funding from the European Union's Horizon 2020 research and innovation program under the ERC grant SUPERGRAPH No. 866365. B.S., H.S. and W.Y. acknowledge support from the QuantERA II Program that has received funding from the European Union's Horizon 2020 research and innovation program under Grant Agreement No 101017733. B.S. acknowledge funding from the ANR agency under the 'France 2030' plan, with Reference No. ANR-22-PETQ-0003.
A.G.G. and S. F. acknowledge financial support from the European Research Council (ERC) Consolidator grant under grant agreement No. 101042707 (TOPOMORPH). 

\section*{Competing Interests} The authors declare that they have no competing interests.

\bibliography{Graphene-QSH}

\begin{thebibliography}{57}%
\makeatletter
\providecommand \@ifxundefined [1]{%
 \@ifx{#1\undefined}
}%
\providecommand \@ifnum [1]{%
 \ifnum #1\expandafter \@firstoftwo
 \else \expandafter \@secondoftwo
 \fi
}%
\providecommand \@ifx [1]{%
 \ifx #1\expandafter \@firstoftwo
 \else \expandafter \@secondoftwo
 \fi
}%
\providecommand \natexlab [1]{#1}%
\providecommand \enquote  [1]{``#1''}%
\providecommand \bibnamefont  [1]{#1}%
\providecommand \bibfnamefont [1]{#1}%
\providecommand \citenamefont [1]{#1}%
\providecommand \href@noop [0]{\@secondoftwo}%
\providecommand \href [0]{\begingroup \@sanitize@url \@href}%
\providecommand \@href[1]{\@@startlink{#1}\@@href}%
\providecommand \@@href[1]{\endgroup#1\@@endlink}%
\providecommand \@sanitize@url [0]{\catcode `\\12\catcode `\$12\catcode
  `\&12\catcode `\#12\catcode `\^12\catcode `\_12\catcode `\%12\relax}%
\providecommand \@@startlink[1]{}%
\providecommand \@@endlink[0]{}%
\providecommand \url  [0]{\begingroup\@sanitize@url \@url }%
\providecommand \@url [1]{\endgroup\@href {#1}{\urlprefix }}%
\providecommand \urlprefix  [0]{URL }%
\providecommand \Eprint [0]{\href }%
\providecommand \doibase [0]{https://doi.org/}%
\providecommand \selectlanguage [0]{\@gobble}%
\providecommand \bibinfo  [0]{\@secondoftwo}%
\providecommand \bibfield  [0]{\@secondoftwo}%
\providecommand \translation [1]{[#1]}%
\providecommand \BibitemOpen [0]{}%
\providecommand \bibitemStop [0]{}%
\providecommand \bibitemNoStop [0]{.\EOS\space}%
\providecommand \EOS [0]{\spacefactor3000\relax}%
\providecommand \BibitemShut  [1]{\csname bibitem#1\endcsname}%
\let\auto@bib@innerbib\@empty
\bibitem [{\citenamefont {Fu}\ and\ \citenamefont {Kane}(2008)}]{Fu08}%
  \BibitemOpen
  \bibfield  {author} {\bibinfo {author} {\bibfnamefont {L.}~\bibnamefont
  {Fu}}\ and\ \bibinfo {author} {\bibfnamefont {C.~L.}\ \bibnamefont {Kane}},\
  }\bibfield  {title} {\bibinfo {title} {{Superconducting Proximity Effect and
  Majorana Fermions at the Surface of a Topological Insulator}},\ }\href
  {https://doi.org/10.1103/PhysRevLett.100.096407} {\bibfield  {journal}
  {\bibinfo  {journal} {Phys. Rev. Lett.}\ }\textbf {\bibinfo {volume} {100}},\
  \bibinfo {pages} {096407} (\bibinfo {year} {2008})}\BibitemShut {NoStop}%
\bibitem [{\citenamefont {Fu}\ and\ \citenamefont {Kane}(2009)}]{Fu09}%
  \BibitemOpen
  \bibfield  {author} {\bibinfo {author} {\bibfnamefont {L.}~\bibnamefont
  {Fu}}\ and\ \bibinfo {author} {\bibfnamefont {C.~L.}\ \bibnamefont {Kane}},\
  }\bibfield  {title} {\bibinfo {title} {{Josephson current and noise at a
  superconductor/quantum-spin-Hall-insulator/superconductor junction}},\ }\href
  {https://doi.org/10.1103/PhysRevB.79.161408} {\bibfield  {journal} {\bibinfo
  {journal} {Phys. Rev. B}\ }\textbf {\bibinfo {volume} {79}},\ \bibinfo
  {pages} {161408} (\bibinfo {year} {2009})}\BibitemShut {NoStop}%
\bibitem [{\citenamefont {Nayak}\ \emph {et~al.}(2008)\citenamefont {Nayak},
  \citenamefont {Simon}, \citenamefont {Stern}, \citenamefont {Freedman},\ and\
  \citenamefont {Das~Sarma}}]{Nayak08}%
  \BibitemOpen
  \bibfield  {author} {\bibinfo {author} {\bibfnamefont {C.}~\bibnamefont
  {Nayak}}, \bibinfo {author} {\bibfnamefont {S.~H.}\ \bibnamefont {Simon}},
  \bibinfo {author} {\bibfnamefont {A.}~\bibnamefont {Stern}}, \bibinfo
  {author} {\bibfnamefont {M.}~\bibnamefont {Freedman}},\ and\ \bibinfo
  {author} {\bibfnamefont {S.}~\bibnamefont {Das~Sarma}},\ }\bibfield  {title}
  {\bibinfo {title} {{Non-Abelian anyons and topological quantum
  computation}},\ }\href {https://doi.org/10.1103/RevModPhys.80.1083}
  {\bibfield  {journal} {\bibinfo  {journal} {Rev. Mod. Phys.}\ }\textbf
  {\bibinfo {volume} {80}},\ \bibinfo {pages} {1083} (\bibinfo {year}
  {2008})}\BibitemShut {NoStop}%
\bibitem [{\citenamefont {Hasan}\ and\ \citenamefont {Kane}(2010)}]{Hasan10}%
  \BibitemOpen
  \bibfield  {author} {\bibinfo {author} {\bibfnamefont {M.~Z.}\ \bibnamefont
  {Hasan}}\ and\ \bibinfo {author} {\bibfnamefont {C.~L.}\ \bibnamefont
  {Kane}},\ }\bibfield  {title} {\bibinfo {title} {{Colloquium: Topological
  insulators}},\ }\href {https://doi.org/10.1103/RevModPhys.82.3045} {\bibfield
   {journal} {\bibinfo  {journal} {Rev. Mod. Phys.}\ }\textbf {\bibinfo
  {volume} {82}},\ \bibinfo {pages} {3045} (\bibinfo {year}
  {2010})}\BibitemShut {NoStop}%
\bibitem [{\citenamefont {Qi}\ and\ \citenamefont {Zhang}(2011)}]{Qi2011}%
  \BibitemOpen
  \bibfield  {author} {\bibinfo {author} {\bibfnamefont {X.-L.}\ \bibnamefont
  {Qi}}\ and\ \bibinfo {author} {\bibfnamefont {S.-C.}\ \bibnamefont {Zhang}},\
  }\bibfield  {title} {\bibinfo {title} {Topological insulators and
  superconductors},\ }\href {https://doi.org/10.1103/RevModPhys.83.1057}
  {\bibfield  {journal} {\bibinfo  {journal} {Rev. Mod. Phys.}\ }\textbf
  {\bibinfo {volume} {83}},\ \bibinfo {pages} {1057} (\bibinfo {year}
  {2011})}\BibitemShut {NoStop}%
\bibitem [{\citenamefont {Yazdani}\ \emph {et~al.}(2023)\citenamefont
  {Yazdani}, \citenamefont {von Oppen}, \citenamefont {Halperin},\ and\
  \citenamefont {Yacoby}}]{Yazdani23}%
  \BibitemOpen
  \bibfield  {author} {\bibinfo {author} {\bibfnamefont {A.}~\bibnamefont
  {Yazdani}}, \bibinfo {author} {\bibfnamefont {F.}~\bibnamefont {von Oppen}},
  \bibinfo {author} {\bibfnamefont {B.~I.}\ \bibnamefont {Halperin}},\ and\
  \bibinfo {author} {\bibfnamefont {A.}~\bibnamefont {Yacoby}},\ }\bibfield
  {title} {\bibinfo {title} {{Hunting for Majoranas}},\ }\bibfield  {journal}
  {\bibinfo  {journal} {Science}\ }\textbf {\bibinfo {volume} {380}},\ \href
  {https://doi.org/10.1126/science.ade0850} {10.1126/science.ade0850} (\bibinfo
  {year} {2023})\BibitemShut {NoStop}%
\bibitem [{\citenamefont {Rokhinson}\ \emph {et~al.}(2012)\citenamefont
  {Rokhinson}, \citenamefont {Liu},\ and\ \citenamefont
  {Furdyna}}]{Rokhinson2012}%
  \BibitemOpen
  \bibfield  {author} {\bibinfo {author} {\bibfnamefont {L.~P.}\ \bibnamefont
  {Rokhinson}}, \bibinfo {author} {\bibfnamefont {X.}~\bibnamefont {Liu}},\
  and\ \bibinfo {author} {\bibfnamefont {J.~K.}\ \bibnamefont {Furdyna}},\
  }\bibfield  {title} {\bibinfo {title} {{The fractional a.c.\ Josephson effect
  in a semiconductor-superconductor nanowire as a signature of Majorana
  particles}},\ }\href {https://doi.org/10.1038/nphys2429} {\bibfield
  {journal} {\bibinfo  {journal} {Nature Physics}\ }\textbf {\bibinfo {volume}
  {8}},\ \bibinfo {pages} {795} (\bibinfo {year} {2012})}\BibitemShut {NoStop}%
\bibitem [{\citenamefont {Wiedenmann}\ \emph {et~al.}(2016)\citenamefont
  {Wiedenmann}, \citenamefont {Bocquillon}, \citenamefont {Deacon},
  \citenamefont {Hartinger}, \citenamefont {Herrmann}, \citenamefont
  {Klapwijk}, \citenamefont {Maier}, \citenamefont {Ames}, \citenamefont
  {Br{\"u}ne}, \citenamefont {Gould}, \citenamefont {Oiwa}, \citenamefont
  {Ishibashi}, \citenamefont {Tarucha}, \citenamefont {Buhmann},\ and\
  \citenamefont {Molenkamp}}]{Wiedenmann2016}%
  \BibitemOpen
  \bibfield  {author} {\bibinfo {author} {\bibfnamefont {J.}~\bibnamefont
  {Wiedenmann}}, \bibinfo {author} {\bibfnamefont {E.}~\bibnamefont
  {Bocquillon}}, \bibinfo {author} {\bibfnamefont {R.~S.}\ \bibnamefont
  {Deacon}}, \bibinfo {author} {\bibfnamefont {S.}~\bibnamefont {Hartinger}},
  \bibinfo {author} {\bibfnamefont {O.}~\bibnamefont {Herrmann}}, \bibinfo
  {author} {\bibfnamefont {T.~M.}\ \bibnamefont {Klapwijk}}, \bibinfo {author}
  {\bibfnamefont {L.}~\bibnamefont {Maier}}, \bibinfo {author} {\bibfnamefont
  {C.}~\bibnamefont {Ames}}, \bibinfo {author} {\bibfnamefont {C.}~\bibnamefont
  {Br{\"u}ne}}, \bibinfo {author} {\bibfnamefont {C.}~\bibnamefont {Gould}},
  \bibinfo {author} {\bibfnamefont {A.}~\bibnamefont {Oiwa}}, \bibinfo {author}
  {\bibfnamefont {K.}~\bibnamefont {Ishibashi}}, \bibinfo {author}
  {\bibfnamefont {S.}~\bibnamefont {Tarucha}}, \bibinfo {author} {\bibfnamefont
  {H.}~\bibnamefont {Buhmann}},\ and\ \bibinfo {author} {\bibfnamefont {L.~W.}\
  \bibnamefont {Molenkamp}},\ }\bibfield  {title} {\bibinfo {title}
  {4$\pi$-periodic {Josephson supercurrent in HgTe-based topological Josephson
  junctions}},\ }\href {https://doi.org/10.1038/ncomms10303} {\bibfield
  {journal} {\bibinfo  {journal} {Nature Communications}\ ,\ \bibinfo {pages}
  {10303}} (\bibinfo {year} {2016})}\BibitemShut {NoStop}%
\bibitem [{\citenamefont {Mourik}\ \emph {et~al.}(2012)\citenamefont {Mourik},
  \citenamefont {Zuo}, \citenamefont {Frolov}, \citenamefont {Plissard},
  \citenamefont {Bakkers},\ and\ \citenamefont {Kouwenhoven}}]{Mourik2012}%
  \BibitemOpen
  \bibfield  {author} {\bibinfo {author} {\bibfnamefont {V.}~\bibnamefont
  {Mourik}}, \bibinfo {author} {\bibfnamefont {K.}~\bibnamefont {Zuo}},
  \bibinfo {author} {\bibfnamefont {S.~M.}\ \bibnamefont {Frolov}}, \bibinfo
  {author} {\bibfnamefont {S.~R.}\ \bibnamefont {Plissard}}, \bibinfo {author}
  {\bibfnamefont {E.~P.~A.~M.}\ \bibnamefont {Bakkers}},\ and\ \bibinfo
  {author} {\bibfnamefont {L.~P.}\ \bibnamefont {Kouwenhoven}},\ }\bibfield
  {title} {\bibinfo {title} {Signatures of {Majorana} fermions in hybrid
  superconductor-semiconductor nanowire devices},\ }\href
  {https://doi.org/10.1126/science.1222360} {\bibfield  {journal} {\bibinfo
  {journal} {Science}\ }\textbf {\bibinfo {volume} {336}},\ \bibinfo {pages}
  {1003} (\bibinfo {year} {2012})}\BibitemShut {NoStop}%
\bibitem [{\citenamefont {Abanin}\ \emph {et~al.}(2006)\citenamefont {Abanin},
  \citenamefont {Lee},\ and\ \citenamefont {Levitov}}]{Abanin06}%
  \BibitemOpen
  \bibfield  {author} {\bibinfo {author} {\bibfnamefont {D.~A.}\ \bibnamefont
  {Abanin}}, \bibinfo {author} {\bibfnamefont {P.~A.}\ \bibnamefont {Lee}},\
  and\ \bibinfo {author} {\bibfnamefont {L.~S.}\ \bibnamefont {Levitov}},\
  }\bibfield  {title} {\bibinfo {title} {{Spin-Filtered Edge States and Quantum
  Hall Effect in Graphene}},\ }\href
  {https://doi.org/10.1103/PhysRevLett.96.176803} {\bibfield  {journal}
  {\bibinfo  {journal} {Phys. Rev. Lett.}\ }\textbf {\bibinfo {volume} {96}},\
  \bibinfo {pages} {176803} (\bibinfo {year} {2006})}\BibitemShut {NoStop}%
\bibitem [{\citenamefont {Fertig}\ and\ \citenamefont {Brey}(2006)}]{Fertig06}%
  \BibitemOpen
  \bibfield  {author} {\bibinfo {author} {\bibfnamefont {H.~A.}\ \bibnamefont
  {Fertig}}\ and\ \bibinfo {author} {\bibfnamefont {L.}~\bibnamefont {Brey}},\
  }\bibfield  {title} {\bibinfo {title} {{Luttinger Liquid at the Edge of
  Undoped Graphene in a Strong Magnetic Field}},\ }\href
  {https://doi.org/10.1103/PhysRevLett.97.116805} {\bibfield  {journal}
  {\bibinfo  {journal} {Phys. Rev. Lett.}\ }\textbf {\bibinfo {volume} {97}},\
  \bibinfo {pages} {116805} (\bibinfo {year} {2006})}\BibitemShut {NoStop}%
\bibitem [{\citenamefont {Kharitonov}\ \emph {et~al.}(2016)\citenamefont
  {Kharitonov}, \citenamefont {Juergens},\ and\ \citenamefont
  {Trauzettel}}]{Kharitonov16}%
  \BibitemOpen
  \bibfield  {author} {\bibinfo {author} {\bibfnamefont {M.}~\bibnamefont
  {Kharitonov}}, \bibinfo {author} {\bibfnamefont {S.}~\bibnamefont
  {Juergens}},\ and\ \bibinfo {author} {\bibfnamefont {B.}~\bibnamefont
  {Trauzettel}},\ }\bibfield  {title} {\bibinfo {title} {{Interplay of topology
  and interactions in quantum Hall topological insulators: U(1) symmetry,
  tunable Luttinger liquid, and interaction-induced phase transitions}},\
  }\href {https://doi.org/10.1103/PhysRevB.94.035146} {\bibfield  {journal}
  {\bibinfo  {journal} {Phys. Rev. B}\ }\textbf {\bibinfo {volume} {94}},\
  \bibinfo {pages} {035146} (\bibinfo {year} {2016})}\BibitemShut {NoStop}%
\bibitem [{\citenamefont {Alicea}\ and\ \citenamefont
  {Fisher}(2006)}]{Alicea06}%
  \BibitemOpen
  \bibfield  {author} {\bibinfo {author} {\bibfnamefont {J.}~\bibnamefont
  {Alicea}}\ and\ \bibinfo {author} {\bibfnamefont {M.~P.~A.}\ \bibnamefont
  {Fisher}},\ }\bibfield  {title} {\bibinfo {title} {Graphene integer quantum
  {Hall} effect in the ferromagnetic and paramagnetic regimes},\ }\href
  {https://doi.org/10.1103/PhysRevB.74.075422} {\bibfield  {journal} {\bibinfo
  {journal} {Phys. Rev. B}\ }\textbf {\bibinfo {volume} {74}},\ \bibinfo
  {pages} {075422} (\bibinfo {year} {2006})}\BibitemShut {NoStop}%
\bibitem [{\citenamefont {Yang}\ \emph {et~al.}(2006)\citenamefont {Yang},
  \citenamefont {Das~Sarma},\ and\ \citenamefont {MacDonald}}]{Yang06}%
  \BibitemOpen
  \bibfield  {author} {\bibinfo {author} {\bibfnamefont {K.}~\bibnamefont
  {Yang}}, \bibinfo {author} {\bibfnamefont {S.}~\bibnamefont {Das~Sarma}},\
  and\ \bibinfo {author} {\bibfnamefont {A.~H.}\ \bibnamefont {MacDonald}},\
  }\bibfield  {title} {\bibinfo {title} {{Collective modes and skyrmion
  excitations in graphene $SU(4)$ quantum Hall ferromagnets}},\ }\href
  {https://doi.org/10.1103/PhysRevB.74.075423} {\bibfield  {journal} {\bibinfo
  {journal} {Phys. Rev. B}\ }\textbf {\bibinfo {volume} {74}},\ \bibinfo
  {pages} {075423} (\bibinfo {year} {2006})}\BibitemShut {NoStop}%
\bibitem [{\citenamefont {Herbut}(2007)}]{Herbut07}%
  \BibitemOpen
  \bibfield  {author} {\bibinfo {author} {\bibfnamefont {I.~F.}\ \bibnamefont
  {Herbut}},\ }\bibfield  {title} {\bibinfo {title} {{Theory of integer quantum
  Hall effect in graphene}},\ }\href
  {https://doi.org/10.1103/PhysRevB.75.165411} {\bibfield  {journal} {\bibinfo
  {journal} {Phys. Rev. B}\ }\textbf {\bibinfo {volume} {75}},\ \bibinfo
  {pages} {165411} (\bibinfo {year} {2007})}\BibitemShut {NoStop}%
\bibitem [{\citenamefont {Kharitonov}(2012)}]{Kharitonov12}%
  \BibitemOpen
  \bibfield  {author} {\bibinfo {author} {\bibfnamefont {M.}~\bibnamefont
  {Kharitonov}},\ }\bibfield  {title} {\bibinfo {title} {{Phase diagram for the
  $\ensuremath{\nu}=0$ quantum Hall state in monolayer graphene}},\ }\href
  {https://doi.org/10.1103/PhysRevB.85.155439} {\bibfield  {journal} {\bibinfo
  {journal} {Phys. Rev. B}\ }\textbf {\bibinfo {volume} {85}},\ \bibinfo
  {pages} {155439} (\bibinfo {year} {2012})}\BibitemShut {NoStop}%
\bibitem [{\citenamefont {Wei}\ \emph {et~al.}(2025)\citenamefont {Wei},
  \citenamefont {Xu}, \citenamefont {Villadiego},\ and\ \citenamefont
  {Huang}}]{Wei25}%
  \BibitemOpen
  \bibfield  {author} {\bibinfo {author} {\bibfnamefont {N.}~\bibnamefont
  {Wei}}, \bibinfo {author} {\bibfnamefont {G.}~\bibnamefont {Xu}}, \bibinfo
  {author} {\bibfnamefont {I.~S.}\ \bibnamefont {Villadiego}},\ and\ \bibinfo
  {author} {\bibfnamefont {C.}~\bibnamefont {Huang}},\ }\bibfield  {title}
  {\bibinfo {title} {{Landau-Level Mixing and $SU(4)$ Symmetry Breaking in
  Graphene}},\ }\href {https://doi.org/10.1103/PhysRevLett.134.046501}
  {\bibfield  {journal} {\bibinfo  {journal} {Phys. Rev. Lett.}\ }\textbf
  {\bibinfo {volume} {134}},\ \bibinfo {pages} {046501} (\bibinfo {year}
  {2025})}\BibitemShut {NoStop}%
\bibitem [{\citenamefont {An}\ and\ \citenamefont {Murthy}(2025)}]{An25}%
  \BibitemOpen
  \bibfield  {author} {\bibinfo {author} {\bibfnamefont {J.}~\bibnamefont
  {An}}\ and\ \bibinfo {author} {\bibfnamefont {G.}~\bibnamefont {Murthy}},\
  }\bibfield  {title} {\bibinfo {title} {{Uniquely identifying quantum Hall
  phases in charge-neutral graphene}},\ }\href
  {https://doi.org/10.1103/PhysRevB.111.195121} {\bibfield  {journal} {\bibinfo
   {journal} {Phys. Rev. B}\ }\textbf {\bibinfo {volume} {111}},\ \bibinfo
  {pages} {195121} (\bibinfo {year} {2025})}\BibitemShut {NoStop}%
\bibitem [{\citenamefont {Young}\ \emph {et~al.}(2014)\citenamefont {Young},
  \citenamefont {Sanchez-Yamagishi}, \citenamefont {Hunt}, \citenamefont
  {Choi}, \citenamefont {Watanabe}, \citenamefont {Taniguchi}, \citenamefont
  {Ashoori},\ and\ \citenamefont {Jarillo-Herrero}}]{Young14}%
  \BibitemOpen
  \bibfield  {author} {\bibinfo {author} {\bibfnamefont {A.~F.}\ \bibnamefont
  {Young}}, \bibinfo {author} {\bibfnamefont {J.~D.}\ \bibnamefont
  {Sanchez-Yamagishi}}, \bibinfo {author} {\bibfnamefont {B.}~\bibnamefont
  {Hunt}}, \bibinfo {author} {\bibfnamefont {S.~H.}\ \bibnamefont {Choi}},
  \bibinfo {author} {\bibfnamefont {K.}~\bibnamefont {Watanabe}}, \bibinfo
  {author} {\bibfnamefont {T.}~\bibnamefont {Taniguchi}}, \bibinfo {author}
  {\bibfnamefont {R.~C.}\ \bibnamefont {Ashoori}},\ and\ \bibinfo {author}
  {\bibfnamefont {P.}~\bibnamefont {Jarillo-Herrero}},\ }\bibfield  {title}
  {\bibinfo {title} {{Tunable symmetry breaking and helical edge transport in a
  graphene quantum spin Hall state}},\ }\href
  {https://doi.org/10.1038/nature12800} {\bibfield  {journal} {\bibinfo
  {journal} {Nature}\ }\textbf {\bibinfo {volume} {505}},\ \bibinfo {pages}
  {528} (\bibinfo {year} {2014})}\BibitemShut {NoStop}%
\bibitem [{\citenamefont {Maher}\ \emph {et~al.}(2013)\citenamefont {Maher},
  \citenamefont {Dean}, \citenamefont {Young}, \citenamefont {Taniguchi},
  \citenamefont {Watanabe}, \citenamefont {Shepard}, \citenamefont {Hone},\
  and\ \citenamefont {Kim}}]{Maher13}%
  \BibitemOpen
  \bibfield  {author} {\bibinfo {author} {\bibfnamefont {P.}~\bibnamefont
  {Maher}}, \bibinfo {author} {\bibfnamefont {C.~R.}\ \bibnamefont {Dean}},
  \bibinfo {author} {\bibfnamefont {A.~F.}\ \bibnamefont {Young}}, \bibinfo
  {author} {\bibfnamefont {T.}~\bibnamefont {Taniguchi}}, \bibinfo {author}
  {\bibfnamefont {K.}~\bibnamefont {Watanabe}}, \bibinfo {author}
  {\bibfnamefont {K.~L.}\ \bibnamefont {Shepard}}, \bibinfo {author}
  {\bibfnamefont {J.}~\bibnamefont {Hone}},\ and\ \bibinfo {author}
  {\bibfnamefont {P.}~\bibnamefont {Kim}},\ }\bibfield  {title} {\bibinfo
  {title} {Evidence for a spin phase transition at charge neutrality in bilayer
  graphene},\ }\href {https://doi.org/10.1038/nphys2528} {\bibfield  {journal}
  {\bibinfo  {journal} {Nature Physics}\ }\textbf {\bibinfo {volume} {9}},\
  \bibinfo {pages} {154} (\bibinfo {year} {2013})}\BibitemShut {NoStop}%
\bibitem [{\citenamefont {Veyrat}\ \emph {et~al.}(2020)\citenamefont {Veyrat},
  \citenamefont {D{\'e}prez}, \citenamefont {Coissard}, \citenamefont {Li},
  \citenamefont {Gay}, \citenamefont {Watanabe}, \citenamefont {Taniguchi},
  \citenamefont {Han}, \citenamefont {Piot}, \citenamefont {Sellier} \emph
  {et~al.}}]{Veyrat20}%
  \BibitemOpen
  \bibfield  {author} {\bibinfo {author} {\bibfnamefont {L.}~\bibnamefont
  {Veyrat}}, \bibinfo {author} {\bibfnamefont {C.}~\bibnamefont {D{\'e}prez}},
  \bibinfo {author} {\bibfnamefont {A.}~\bibnamefont {Coissard}}, \bibinfo
  {author} {\bibfnamefont {X.}~\bibnamefont {Li}}, \bibinfo {author}
  {\bibfnamefont {F.}~\bibnamefont {Gay}}, \bibinfo {author} {\bibfnamefont
  {K.}~\bibnamefont {Watanabe}}, \bibinfo {author} {\bibfnamefont
  {T.}~\bibnamefont {Taniguchi}}, \bibinfo {author} {\bibfnamefont
  {Z.}~\bibnamefont {Han}}, \bibinfo {author} {\bibfnamefont {B.~A.}\
  \bibnamefont {Piot}}, \bibinfo {author} {\bibfnamefont {H.}~\bibnamefont
  {Sellier}}, \emph {et~al.},\ }\bibfield  {title} {\bibinfo {title} {{Helical
  quantum Hall phase in graphene on $\rm{SrTiO}_3$}},\ }\href
  {https://doi.org/10.1126/science.aax8201} {\bibfield  {journal} {\bibinfo
  {journal} {Science}\ }\textbf {\bibinfo {volume} {367}},\ \bibinfo {pages}
  {781} (\bibinfo {year} {2020})}\BibitemShut {NoStop}%
\bibitem [{\citenamefont {Coissard}\ \emph {et~al.}(2022)\citenamefont
  {Coissard}, \citenamefont {Wander}, \citenamefont {Vignaud}, \citenamefont
  {Grushin}, \citenamefont {Repellin}, \citenamefont {Watanabe}, \citenamefont
  {Taniguchi}, \citenamefont {Gay}, \citenamefont {Winkelmann}, \citenamefont
  {Courtois}, \citenamefont {Sellier},\ and\ \citenamefont
  {Sac{\'e}p{\'e}}}]{coissard2022a}%
  \BibitemOpen
  \bibfield  {author} {\bibinfo {author} {\bibfnamefont {A.}~\bibnamefont
  {Coissard}}, \bibinfo {author} {\bibfnamefont {D.}~\bibnamefont {Wander}},
  \bibinfo {author} {\bibfnamefont {H.}~\bibnamefont {Vignaud}}, \bibinfo
  {author} {\bibfnamefont {A.~G.}\ \bibnamefont {Grushin}}, \bibinfo {author}
  {\bibfnamefont {C.}~\bibnamefont {Repellin}}, \bibinfo {author}
  {\bibfnamefont {K.}~\bibnamefont {Watanabe}}, \bibinfo {author}
  {\bibfnamefont {T.}~\bibnamefont {Taniguchi}}, \bibinfo {author}
  {\bibfnamefont {F.}~\bibnamefont {Gay}}, \bibinfo {author} {\bibfnamefont
  {C.~B.}\ \bibnamefont {Winkelmann}}, \bibinfo {author} {\bibfnamefont
  {H.}~\bibnamefont {Courtois}}, \bibinfo {author} {\bibfnamefont
  {H.}~\bibnamefont {Sellier}},\ and\ \bibinfo {author} {\bibfnamefont
  {B.}~\bibnamefont {Sac{\'e}p{\'e}}},\ }\bibfield  {title} {\bibinfo {title}
  {Imaging tunable quantum {Hall} broken-symmetry orders in graphene},\ }\href
  {https://doi.org/10.1038/s41586-022-04513-7} {\bibfield  {journal} {\bibinfo
  {journal} {Nature}\ }\textbf {\bibinfo {volume} {605}},\ \bibinfo {pages}
  {51} (\bibinfo {year} {2022})}\BibitemShut {NoStop}%
\bibitem [{\citenamefont {Amet}\ \emph {et~al.}(2016)\citenamefont {Amet},
  \citenamefont {Ke}, \citenamefont {Borzenets}, \citenamefont {Wang},
  \citenamefont {Watanabe}, \citenamefont {Taniguchi}, \citenamefont {Deacon},
  \citenamefont {Yamamoto}, \citenamefont {Bomze},\ and\ \citenamefont
  {Finkelstein}}]{Amet16}%
  \BibitemOpen
  \bibfield  {author} {\bibinfo {author} {\bibfnamefont {F.}~\bibnamefont
  {Amet}}, \bibinfo {author} {\bibfnamefont {C.~T.}\ \bibnamefont {Ke}},
  \bibinfo {author} {\bibfnamefont {I.~V.}\ \bibnamefont {Borzenets}}, \bibinfo
  {author} {\bibfnamefont {J.}~\bibnamefont {Wang}}, \bibinfo {author}
  {\bibfnamefont {K.}~\bibnamefont {Watanabe}}, \bibinfo {author}
  {\bibfnamefont {T.}~\bibnamefont {Taniguchi}}, \bibinfo {author}
  {\bibfnamefont {R.~S.}\ \bibnamefont {Deacon}}, \bibinfo {author}
  {\bibfnamefont {M.}~\bibnamefont {Yamamoto}}, \bibinfo {author}
  {\bibfnamefont {S.}~\bibnamefont {Bomze}, \bibfnamefont {Y.~Tarucha}},\ and\
  \bibinfo {author} {\bibfnamefont {G.}~\bibnamefont {Finkelstein}},\
  }\bibfield  {title} {\bibinfo {title} {{Supercurrent in the quantum Hall
  regime}},\ }\href {https://doi.org/10.1126/science.aad6203} {\bibfield
  {journal} {\bibinfo  {journal} {Science}\ }\textbf {\bibinfo {volume}
  {352}},\ \bibinfo {pages} {966} (\bibinfo {year} {2016})}\BibitemShut
  {NoStop}%
\bibitem [{\citenamefont {Lee}\ \emph {et~al.}(2017)\citenamefont {Lee},
  \citenamefont {Huang}, \citenamefont {Efetov}, \citenamefont {Wei},
  \citenamefont {Hart}, \citenamefont {Taniguchi}, \citenamefont {Watanabe},
  \citenamefont {Yacoby},\ and\ \citenamefont {Kim}}]{Lee2017}%
  \BibitemOpen
  \bibfield  {author} {\bibinfo {author} {\bibfnamefont {G.-H.}\ \bibnamefont
  {Lee}}, \bibinfo {author} {\bibfnamefont {K.-F.}\ \bibnamefont {Huang}},
  \bibinfo {author} {\bibfnamefont {D.~K.}\ \bibnamefont {Efetov}}, \bibinfo
  {author} {\bibfnamefont {D.~S.}\ \bibnamefont {Wei}}, \bibinfo {author}
  {\bibfnamefont {S.}~\bibnamefont {Hart}}, \bibinfo {author} {\bibfnamefont
  {T.}~\bibnamefont {Taniguchi}}, \bibinfo {author} {\bibfnamefont
  {K.}~\bibnamefont {Watanabe}}, \bibinfo {author} {\bibfnamefont
  {A.}~\bibnamefont {Yacoby}},\ and\ \bibinfo {author} {\bibfnamefont
  {P.}~\bibnamefont {Kim}},\ }\bibfield  {title} {\bibinfo {title} {{Inducing
  superconducting correlation in quantum Hall edge states}},\ }\href
  {https://doi.org/10.1038/nphys4084} {\bibfield  {journal} {\bibinfo
  {journal} {Nature Physics}\ }\textbf {\bibinfo {volume} {13}},\ \bibinfo
  {pages} {693} (\bibinfo {year} {2017})}\BibitemShut {NoStop}%
\bibitem [{\citenamefont {Zhao}\ \emph {et~al.}(2020)\citenamefont {Zhao},
  \citenamefont {Arnault}, \citenamefont {Bondarev}, \citenamefont
  {Seredinski}, \citenamefont {Larson}, \citenamefont {Draelos}, \citenamefont
  {Li}, \citenamefont {Watanabe}, \citenamefont {Taniguchi}, \citenamefont
  {Amet}, \citenamefont {Baranger},\ and\ \citenamefont
  {Finkelstein}}]{Zhao2020a}%
  \BibitemOpen
  \bibfield  {author} {\bibinfo {author} {\bibfnamefont {L.}~\bibnamefont
  {Zhao}}, \bibinfo {author} {\bibfnamefont {E.~G.}\ \bibnamefont {Arnault}},
  \bibinfo {author} {\bibfnamefont {A.}~\bibnamefont {Bondarev}}, \bibinfo
  {author} {\bibfnamefont {A.}~\bibnamefont {Seredinski}}, \bibinfo {author}
  {\bibfnamefont {T.~F.~Q.}\ \bibnamefont {Larson}}, \bibinfo {author}
  {\bibfnamefont {A.~W.}\ \bibnamefont {Draelos}}, \bibinfo {author}
  {\bibfnamefont {H.}~\bibnamefont {Li}}, \bibinfo {author} {\bibfnamefont
  {K.}~\bibnamefont {Watanabe}}, \bibinfo {author} {\bibfnamefont
  {T.}~\bibnamefont {Taniguchi}}, \bibinfo {author} {\bibfnamefont
  {F.}~\bibnamefont {Amet}}, \bibinfo {author} {\bibfnamefont {H.~U.}\
  \bibnamefont {Baranger}},\ and\ \bibinfo {author} {\bibfnamefont
  {G.}~\bibnamefont {Finkelstein}},\ }\bibfield  {title} {\bibinfo {title}
  {{Interference of chiral Andreev edge states}},\ }\href
  {https://doi.org/10.1038/s41567-020-0898-5} {\bibfield  {journal} {\bibinfo
  {journal} {Nature Physics}\ }\textbf {\bibinfo {volume} {16}},\ \bibinfo
  {pages} {862} (\bibinfo {year} {2020})}\BibitemShut {NoStop}%
\bibitem [{\citenamefont {G{\"{u}}l}\ \emph {et~al.}(2022)\citenamefont
  {G{\"{u}}l}, \citenamefont {Ronen}, \citenamefont {Lee}, \citenamefont
  {Shapourian}, \citenamefont {Zauberman}, \citenamefont {H.}, \citenamefont
  {Watanabe}, \citenamefont {Taniguchi}, \citenamefont {Vishwanath},
  \citenamefont {Yacoby},\ and\ \citenamefont {Kim}}]{Gul2022}%
  \BibitemOpen
  \bibfield  {author} {\bibinfo {author} {\bibfnamefont {{\"{O}}.}~\bibnamefont
  {G{\"{u}}l}}, \bibinfo {author} {\bibfnamefont {Y.}~\bibnamefont {Ronen}},
  \bibinfo {author} {\bibfnamefont {S.~Y.}\ \bibnamefont {Lee}}, \bibinfo
  {author} {\bibfnamefont {H.}~\bibnamefont {Shapourian}}, \bibinfo {author}
  {\bibfnamefont {J.}~\bibnamefont {Zauberman}}, \bibinfo {author}
  {\bibfnamefont {L.~Y.}\ \bibnamefont {H.}}, \bibinfo {author} {\bibfnamefont
  {K.}~\bibnamefont {Watanabe}}, \bibinfo {author} {\bibfnamefont
  {T.}~\bibnamefont {Taniguchi}}, \bibinfo {author} {\bibfnamefont
  {A.}~\bibnamefont {Vishwanath}}, \bibinfo {author} {\bibfnamefont
  {A.}~\bibnamefont {Yacoby}},\ and\ \bibinfo {author} {\bibfnamefont
  {P.}~\bibnamefont {Kim}},\ }\bibfield  {title} {\bibinfo {title} {{Andreev
  reflection in the fractional quantum Hall state}},\ }\href
  {https://doi.org/10.1103/PhysRevX.12.021057} {\bibfield  {journal} {\bibinfo
  {journal} {Physical Review X}\ }\textbf {\bibinfo {volume} {12}},\ \bibinfo
  {pages} {021057} (\bibinfo {year} {2022})}\BibitemShut {NoStop}%
\bibitem [{\citenamefont {Vignaud}\ \emph {et~al.}(2023)\citenamefont
  {Vignaud}, \citenamefont {Perconte}, \citenamefont {Yang}, \citenamefont
  {Kousar}, \citenamefont {Wagner}, \citenamefont {Gay}, \citenamefont
  {Watanabe}, \citenamefont {Taniguchi}, \citenamefont {Courtois},
  \citenamefont {Han}, \citenamefont {Sellier},\ and\ \citenamefont
  {Sac{\'e}p{\'e}}}]{Vignaud23}%
  \BibitemOpen
  \bibfield  {author} {\bibinfo {author} {\bibfnamefont {H.}~\bibnamefont
  {Vignaud}}, \bibinfo {author} {\bibfnamefont {D.}~\bibnamefont {Perconte}},
  \bibinfo {author} {\bibfnamefont {W.}~\bibnamefont {Yang}}, \bibinfo {author}
  {\bibfnamefont {B.}~\bibnamefont {Kousar}}, \bibinfo {author} {\bibfnamefont
  {E.}~\bibnamefont {Wagner}}, \bibinfo {author} {\bibfnamefont
  {F.}~\bibnamefont {Gay}}, \bibinfo {author} {\bibfnamefont {K.}~\bibnamefont
  {Watanabe}}, \bibinfo {author} {\bibfnamefont {T.}~\bibnamefont {Taniguchi}},
  \bibinfo {author} {\bibfnamefont {H.}~\bibnamefont {Courtois}}, \bibinfo
  {author} {\bibfnamefont {Z.}~\bibnamefont {Han}}, \bibinfo {author}
  {\bibfnamefont {H.}~\bibnamefont {Sellier}},\ and\ \bibinfo {author}
  {\bibfnamefont {B.}~\bibnamefont {Sac{\'e}p{\'e}}},\ }\bibfield  {title}
  {\bibinfo {title} {{Evidence for chiral supercurrent in quantum Hall
  Josephson junctions}},\ }\href {https://doi.org/10.1038/s41586-023-06764-4}
  {\bibfield  {journal} {\bibinfo  {journal} {Nature}\ }\textbf {\bibinfo
  {volume} {624}},\ \bibinfo {pages} {545} (\bibinfo {year}
  {2023})}\BibitemShut {NoStop}%
\bibitem [{\citenamefont {Wei}\ \emph {et~al.}(2017)\citenamefont {Wei},
  \citenamefont {van~der Sar}, \citenamefont {Sanchez-Yamagishi}, \citenamefont
  {Watanabe}, \citenamefont {Taniguchi}, \citenamefont {Jarillo-Herrero},
  \citenamefont {Halperin},\ and\ \citenamefont {Yacoby}}]{wei2017}%
  \BibitemOpen
  \bibfield  {author} {\bibinfo {author} {\bibfnamefont {D.~S.}\ \bibnamefont
  {Wei}}, \bibinfo {author} {\bibfnamefont {T.}~\bibnamefont {van~der Sar}},
  \bibinfo {author} {\bibfnamefont {J.~D.}\ \bibnamefont {Sanchez-Yamagishi}},
  \bibinfo {author} {\bibfnamefont {K.}~\bibnamefont {Watanabe}}, \bibinfo
  {author} {\bibfnamefont {T.}~\bibnamefont {Taniguchi}}, \bibinfo {author}
  {\bibfnamefont {P.}~\bibnamefont {Jarillo-Herrero}}, \bibinfo {author}
  {\bibfnamefont {B.~I.}\ \bibnamefont {Halperin}},\ and\ \bibinfo {author}
  {\bibfnamefont {A.}~\bibnamefont {Yacoby}},\ }\bibfield  {title} {\bibinfo
  {title} {Mach-zehnder interferometry using spin-and valley-polarized quantum
  hall edge states in graphene},\ }\href
  {https://doi.org/10.1126/sciadv.1700600} {\bibfield  {journal} {\bibinfo
  {journal} {Science Advances}\ }\textbf {\bibinfo {volume} {3}},\ \bibinfo
  {pages} {e1700600} (\bibinfo {year} {2017})}\BibitemShut {NoStop}%
\bibitem [{\citenamefont {Jo}\ \emph {et~al.}(2021)\citenamefont {Jo},
  \citenamefont {Brasseur}, \citenamefont {Assouline}, \citenamefont {Fleury},
  \citenamefont {Sim}, \citenamefont {Watanabe}, \citenamefont {Taniguchi},
  \citenamefont {Dumnernpanich}, \citenamefont {Roche}, \citenamefont
  {Glattli}, \citenamefont {Kumada}, \citenamefont {Parmentier},\ and\
  \citenamefont {Roulleau}}]{Jo21}%
  \BibitemOpen
  \bibfield  {author} {\bibinfo {author} {\bibfnamefont {M.}~\bibnamefont
  {Jo}}, \bibinfo {author} {\bibfnamefont {P.}~\bibnamefont {Brasseur}},
  \bibinfo {author} {\bibfnamefont {A.}~\bibnamefont {Assouline}}, \bibinfo
  {author} {\bibfnamefont {G.}~\bibnamefont {Fleury}}, \bibinfo {author}
  {\bibfnamefont {H.-S.}\ \bibnamefont {Sim}}, \bibinfo {author} {\bibfnamefont
  {K.}~\bibnamefont {Watanabe}}, \bibinfo {author} {\bibfnamefont
  {T.}~\bibnamefont {Taniguchi}}, \bibinfo {author} {\bibfnamefont
  {W.}~\bibnamefont {Dumnernpanich}}, \bibinfo {author} {\bibfnamefont
  {P.}~\bibnamefont {Roche}}, \bibinfo {author} {\bibfnamefont {D.~C.}\
  \bibnamefont {Glattli}}, \bibinfo {author} {\bibfnamefont {N.}~\bibnamefont
  {Kumada}}, \bibinfo {author} {\bibfnamefont {F.~D.}\ \bibnamefont
  {Parmentier}},\ and\ \bibinfo {author} {\bibfnamefont {P.}~\bibnamefont
  {Roulleau}},\ }\bibfield  {title} {\bibinfo {title} {Quantum hall valley
  splitters and a tunable mach-zehnder interferometer in graphene},\ }\href
  {https://doi.org/10.1103/PhysRevLett.126.146803} {\bibfield  {journal}
  {\bibinfo  {journal} {Phys. Rev. Lett.}\ }\textbf {\bibinfo {volume} {126}},\
  \bibinfo {pages} {146803} (\bibinfo {year} {2021})}\BibitemShut {NoStop}%
\bibitem [{\citenamefont {Blasi}\ \emph {et~al.}(2019)\citenamefont {Blasi},
  \citenamefont {Taddei}, \citenamefont {Giovannetti},\ and\ \citenamefont
  {Braggio}}]{Blasi19}%
  \BibitemOpen
  \bibfield  {author} {\bibinfo {author} {\bibfnamefont {G.}~\bibnamefont
  {Blasi}}, \bibinfo {author} {\bibfnamefont {F.}~\bibnamefont {Taddei}},
  \bibinfo {author} {\bibfnamefont {V.}~\bibnamefont {Giovannetti}},\ and\
  \bibinfo {author} {\bibfnamefont {A.}~\bibnamefont {Braggio}},\ }\bibfield
  {title} {\bibinfo {title} {Manipulation of cooper pair entanglement in hybrid
  topological josephson junctions},\ }\href
  {https://doi.org/10.1103/PhysRevB.99.064514} {\bibfield  {journal} {\bibinfo
  {journal} {Phys. Rev. B}\ }\textbf {\bibinfo {volume} {99}},\ \bibinfo
  {pages} {064514} (\bibinfo {year} {2019})}\BibitemShut {NoStop}%
\bibitem [{\citenamefont {Blasi}\ \emph {et~al.}(2023)\citenamefont {Blasi},
  \citenamefont {Haack}, \citenamefont {Giovannetti}, \citenamefont {Taddei},\
  and\ \citenamefont {Braggio}}]{Blasi23}%
  \BibitemOpen
  \bibfield  {author} {\bibinfo {author} {\bibfnamefont {G.}~\bibnamefont
  {Blasi}}, \bibinfo {author} {\bibfnamefont {G.}~\bibnamefont {Haack}},
  \bibinfo {author} {\bibfnamefont {V.}~\bibnamefont {Giovannetti}}, \bibinfo
  {author} {\bibfnamefont {F.}~\bibnamefont {Taddei}},\ and\ \bibinfo {author}
  {\bibfnamefont {A.}~\bibnamefont {Braggio}},\ }\bibfield  {title} {\bibinfo
  {title} {Topological josephson junctions in the integer quantum hall
  regime},\ }\href {https://doi.org/10.1103/PhysRevResearch.5.033142}
  {\bibfield  {journal} {\bibinfo  {journal} {Phys. Rev. Res.}\ }\textbf
  {\bibinfo {volume} {5}},\ \bibinfo {pages} {033142} (\bibinfo {year}
  {2023})}\BibitemShut {NoStop}%
\bibitem [{\citenamefont {Couto}\ \emph {et~al.}(2011)\citenamefont {Couto},
  \citenamefont {Sac\'ep\'e},\ and\ \citenamefont {Morpurgo}}]{Couto11}%
  \BibitemOpen
  \bibfield  {author} {\bibinfo {author} {\bibfnamefont {N.~J.~G.}\
  \bibnamefont {Couto}}, \bibinfo {author} {\bibfnamefont {B.}~\bibnamefont
  {Sac\'ep\'e}},\ and\ \bibinfo {author} {\bibfnamefont {A.~F.}\ \bibnamefont
  {Morpurgo}},\ }\bibfield  {title} {\bibinfo {title} {{Transport through
  graphene on ${\mathrm{SrTiO}}_{3}$}},\ }\href
  {https://doi.org/10.1103/PhysRevLett.107.225501} {\bibfield  {journal}
  {\bibinfo  {journal} {Phys. Rev. Lett.}\ }\textbf {\bibinfo {volume} {107}},\
  \bibinfo {pages} {225501} (\bibinfo {year} {2011})}\BibitemShut {NoStop}%
\bibitem [{\citenamefont {Frenkel}\ \emph {et~al.}(2017)\citenamefont
  {Frenkel}, \citenamefont {Haham}, \citenamefont {Shperber}, \citenamefont
  {Bell}, \citenamefont {Xie}, \citenamefont {Chen}, \citenamefont {Hikita},
  \citenamefont {Hwang}, \citenamefont {Salje},\ and\ \citenamefont
  {Kalisky}}]{Frenkel17}%
  \BibitemOpen
  \bibfield  {author} {\bibinfo {author} {\bibfnamefont {Y.}~\bibnamefont
  {Frenkel}}, \bibinfo {author} {\bibfnamefont {N.}~\bibnamefont {Haham}},
  \bibinfo {author} {\bibfnamefont {Y.}~\bibnamefont {Shperber}}, \bibinfo
  {author} {\bibfnamefont {C.}~\bibnamefont {Bell}}, \bibinfo {author}
  {\bibfnamefont {Y.}~\bibnamefont {Xie}}, \bibinfo {author} {\bibfnamefont
  {Z.}~\bibnamefont {Chen}}, \bibinfo {author} {\bibfnamefont {Y.}~\bibnamefont
  {Hikita}}, \bibinfo {author} {\bibfnamefont {H.~Y.}\ \bibnamefont {Hwang}},
  \bibinfo {author} {\bibfnamefont {E.~K.}\ \bibnamefont {Salje}},\ and\
  \bibinfo {author} {\bibfnamefont {B.}~\bibnamefont {Kalisky}},\ }\bibfield
  {title} {\bibinfo {title} {{Imaging and tuning polarity at $\rm{SrTiO}_3$
  domain walls}},\ }\href {https://doi.org/10.1038/nmat4966} {\bibfield
  {journal} {\bibinfo  {journal} {Nature materials}\ }\textbf {\bibinfo
  {volume} {16}},\ \bibinfo {pages} {1203} (\bibinfo {year}
  {2017})}\BibitemShut {NoStop}%
\bibitem [{\citenamefont {Csonka}(2025)}]{Csonca25}%
  \BibitemOpen
  \bibfield  {author} {\bibinfo {author} {\bibfnamefont {S.}~\bibnamefont
  {Csonka}},\ }\href@noop {} {} (\bibinfo {year} {2025}),\ \bibinfo {note}
  {private communication}\BibitemShut {NoStop}%
\bibitem [{\citenamefont {K{\"o}nig}\ \emph {et~al.}(2007)\citenamefont
  {K{\"o}nig}, \citenamefont {Wiedmann}, \citenamefont {Br{\"u}ne},
  \citenamefont {Roth}, \citenamefont {Buhmann}, \citenamefont {Molenkamp},
  \citenamefont {Qi},\ and\ \citenamefont {Zhang}}]{konig07}%
  \BibitemOpen
  \bibfield  {author} {\bibinfo {author} {\bibfnamefont {M.}~\bibnamefont
  {K{\"o}nig}}, \bibinfo {author} {\bibfnamefont {S.}~\bibnamefont {Wiedmann}},
  \bibinfo {author} {\bibfnamefont {C.}~\bibnamefont {Br{\"u}ne}}, \bibinfo
  {author} {\bibfnamefont {A.}~\bibnamefont {Roth}}, \bibinfo {author}
  {\bibfnamefont {H.}~\bibnamefont {Buhmann}}, \bibinfo {author} {\bibfnamefont
  {L.~W.}\ \bibnamefont {Molenkamp}}, \bibinfo {author} {\bibfnamefont {X.-L.}\
  \bibnamefont {Qi}},\ and\ \bibinfo {author} {\bibfnamefont {S.-C.}\
  \bibnamefont {Zhang}},\ }\bibfield  {title} {\bibinfo {title} {{Quantum Spin
  Hall Insulator State in HgTe Quantum Wells}},\ }\href
  {https://doi.org/10.1126/science.1148047} {\bibfield  {journal} {\bibinfo
  {journal} {Science}\ }\textbf {\bibinfo {volume} {318}},\ \bibinfo {pages}
  {766} (\bibinfo {year} {2007})}\BibitemShut {NoStop}%
\bibitem [{\citenamefont {San-Jose}\ \emph {et~al.}(2015)\citenamefont
  {San-Jose}, \citenamefont {Lado}, \citenamefont {Aguado}, \citenamefont
  {Guinea},\ and\ \citenamefont {Fern\'andez-Rossier}}]{SanJose15}%
  \BibitemOpen
  \bibfield  {author} {\bibinfo {author} {\bibfnamefont {P.}~\bibnamefont
  {San-Jose}}, \bibinfo {author} {\bibfnamefont {J.~L.}\ \bibnamefont {Lado}},
  \bibinfo {author} {\bibfnamefont {R.}~\bibnamefont {Aguado}}, \bibinfo
  {author} {\bibfnamefont {F.}~\bibnamefont {Guinea}},\ and\ \bibinfo {author}
  {\bibfnamefont {J.}~\bibnamefont {Fern\'andez-Rossier}},\ }\bibfield  {title}
  {\bibinfo {title} {{Majorana Zero Modes in Graphene}},\ }\href
  {https://doi.org/10.1103/PhysRevX.5.041042} {\bibfield  {journal} {\bibinfo
  {journal} {Phys. Rev. X}\ }\textbf {\bibinfo {volume} {5}},\ \bibinfo {pages}
  {041042} (\bibinfo {year} {2015})}\BibitemShut {NoStop}%
\bibitem [{\citenamefont {Williams}\ \emph {et~al.}(2007)\citenamefont
  {Williams}, \citenamefont {DiCarlo},\ and\ \citenamefont
  {Marcus}}]{Williams07}%
  \BibitemOpen
  \bibfield  {author} {\bibinfo {author} {\bibfnamefont {J.~R.}\ \bibnamefont
  {Williams}}, \bibinfo {author} {\bibfnamefont {L.}~\bibnamefont {DiCarlo}},\
  and\ \bibinfo {author} {\bibfnamefont {C.~M.}\ \bibnamefont {Marcus}},\
  }\bibfield  {title} {\bibinfo {title} {{Quantum Hall Effect in a
  Gate-Controlled p-n Junction of Graphene}},\ }\href
  {https://doi.org/10.1126/science.1144657} {\bibfield  {journal} {\bibinfo
  {journal} {Science}\ }\textbf {\bibinfo {volume} {317}},\ \bibinfo {pages}
  {638} (\bibinfo {year} {2007})}\BibitemShut {NoStop}%
\bibitem [{\citenamefont {Amet}\ \emph {et~al.}(2014)\citenamefont {Amet},
  \citenamefont {Williams}, \citenamefont {Watanabe}, \citenamefont
  {Taniguchi},\ and\ \citenamefont {Goldhaber-Gordon}}]{Amet13}%
  \BibitemOpen
  \bibfield  {author} {\bibinfo {author} {\bibfnamefont {F.}~\bibnamefont
  {Amet}}, \bibinfo {author} {\bibfnamefont {J.~R.}\ \bibnamefont {Williams}},
  \bibinfo {author} {\bibfnamefont {K.}~\bibnamefont {Watanabe}}, \bibinfo
  {author} {\bibfnamefont {T.}~\bibnamefont {Taniguchi}},\ and\ \bibinfo
  {author} {\bibfnamefont {D.}~\bibnamefont {Goldhaber-Gordon}},\ }\bibfield
  {title} {\bibinfo {title} {{Selective Equilibration of Spin-Polarized Quantum
  Hall Edge States in Graphene}},\ }\href
  {https://doi.org/10.1103/PhysRevLett.112.196601} {\bibfield  {journal}
  {\bibinfo  {journal} {Phys. Rev. Lett.}\ }\textbf {\bibinfo {volume} {112}},\
  \bibinfo {pages} {196601} (\bibinfo {year} {2014})}\BibitemShut {NoStop}%
\bibitem [{\citenamefont {Roth}\ \emph {et~al.}(2009)\citenamefont {Roth},
  \citenamefont {Br{\"u}ne}, \citenamefont {Buhmann}, \citenamefont
  {Molenkamp}, \citenamefont {Maciejko}, \citenamefont {Qi},\ and\
  \citenamefont {Zhang}}]{Roth09}%
  \BibitemOpen
  \bibfield  {author} {\bibinfo {author} {\bibfnamefont {A.}~\bibnamefont
  {Roth}}, \bibinfo {author} {\bibfnamefont {C.}~\bibnamefont {Br{\"u}ne}},
  \bibinfo {author} {\bibfnamefont {H.}~\bibnamefont {Buhmann}}, \bibinfo
  {author} {\bibfnamefont {L.~W.}\ \bibnamefont {Molenkamp}}, \bibinfo {author}
  {\bibfnamefont {J.}~\bibnamefont {Maciejko}}, \bibinfo {author}
  {\bibfnamefont {X.-L.}\ \bibnamefont {Qi}},\ and\ \bibinfo {author}
  {\bibfnamefont {S.-C.}\ \bibnamefont {Zhang}},\ }\bibfield  {title} {\bibinfo
  {title} {{Nonlocal Transport in the Quantum Spin Hall State}},\ }\href
  {https://doi.org/10.1126/science.1174736} {\bibfield  {journal} {\bibinfo
  {journal} {Science}\ }\textbf {\bibinfo {volume} {325}},\ \bibinfo {pages}
  {294} (\bibinfo {year} {2009})}\BibitemShut {NoStop}%
\bibitem [{\citenamefont {Fei}\ \emph {et~al.}(2017)\citenamefont {Fei},
  \citenamefont {Palomaki}, \citenamefont {Wu}, \citenamefont {Zhao},
  \citenamefont {Cai}, \citenamefont {Sun}, \citenamefont {Nguyen},
  \citenamefont {Finney}, \citenamefont {Xu},\ and\ \citenamefont
  {Cobden}}]{Fei17}%
  \BibitemOpen
  \bibfield  {author} {\bibinfo {author} {\bibfnamefont {Z.}~\bibnamefont
  {Fei}}, \bibinfo {author} {\bibfnamefont {T.}~\bibnamefont {Palomaki}},
  \bibinfo {author} {\bibfnamefont {S.}~\bibnamefont {Wu}}, \bibinfo {author}
  {\bibfnamefont {W.}~\bibnamefont {Zhao}}, \bibinfo {author} {\bibfnamefont
  {X.}~\bibnamefont {Cai}}, \bibinfo {author} {\bibfnamefont {B.}~\bibnamefont
  {Sun}}, \bibinfo {author} {\bibfnamefont {P.}~\bibnamefont {Nguyen}},
  \bibinfo {author} {\bibfnamefont {J.}~\bibnamefont {Finney}}, \bibinfo
  {author} {\bibfnamefont {X.}~\bibnamefont {Xu}},\ and\ \bibinfo {author}
  {\bibfnamefont {D.}~\bibnamefont {Cobden}},\ }\bibfield  {title} {\bibinfo
  {title} {{Edge conduction in monolayer WTe 2}},\ }\href
  {https://doi.org/10.1038/nphys4091} {\bibfield  {journal} {\bibinfo
  {journal} {Nature Physics}\ }\textbf {\bibinfo {volume} {13}},\ \bibinfo
  {pages} {677} (\bibinfo {year} {2017})}\BibitemShut {NoStop}%
\bibitem [{\citenamefont {Wu}\ \emph {et~al.}(2018)\citenamefont {Wu},
  \citenamefont {Fatemi}, \citenamefont {Gibson}, \citenamefont {Watanabe},
  \citenamefont {Taniguchi}, \citenamefont {Cava},\ and\ \citenamefont
  {Jarillo-Herrero}}]{wu18}%
  \BibitemOpen
  \bibfield  {author} {\bibinfo {author} {\bibfnamefont {S.}~\bibnamefont
  {Wu}}, \bibinfo {author} {\bibfnamefont {V.}~\bibnamefont {Fatemi}}, \bibinfo
  {author} {\bibfnamefont {Q.~D.}\ \bibnamefont {Gibson}}, \bibinfo {author}
  {\bibfnamefont {K.}~\bibnamefont {Watanabe}}, \bibinfo {author}
  {\bibfnamefont {T.}~\bibnamefont {Taniguchi}}, \bibinfo {author}
  {\bibfnamefont {R.~J.}\ \bibnamefont {Cava}},\ and\ \bibinfo {author}
  {\bibfnamefont {P.}~\bibnamefont {Jarillo-Herrero}},\ }\bibfield  {title}
  {\bibinfo {title} {{Observation of the quantum spin Hall effect up to 100
  kelvin in a monolayer crystal}},\ }\href
  {https://doi.org/10.1126/science.aan6003} {\bibfield  {journal} {\bibinfo
  {journal} {Science}\ }\textbf {\bibinfo {volume} {359}},\ \bibinfo {pages}
  {76} (\bibinfo {year} {2018})}\BibitemShut {NoStop}%
\bibitem [{\citenamefont {Kang}\ \emph {et~al.}(2024)\citenamefont {Kang},
  \citenamefont {Shen}, \citenamefont {Qiu}, \citenamefont {Zeng},
  \citenamefont {Xia}, \citenamefont {Watanabe}, \citenamefont {Taniguchi},
  \citenamefont {Shan},\ and\ \citenamefont {Mak}}]{Kang2024}%
  \BibitemOpen
  \bibfield  {author} {\bibinfo {author} {\bibfnamefont {K.}~\bibnamefont
  {Kang}}, \bibinfo {author} {\bibfnamefont {B.}~\bibnamefont {Shen}}, \bibinfo
  {author} {\bibfnamefont {Y.}~\bibnamefont {Qiu}}, \bibinfo {author}
  {\bibfnamefont {Y.}~\bibnamefont {Zeng}}, \bibinfo {author} {\bibfnamefont
  {Z.}~\bibnamefont {Xia}}, \bibinfo {author} {\bibfnamefont {K.}~\bibnamefont
  {Watanabe}}, \bibinfo {author} {\bibfnamefont {T.}~\bibnamefont {Taniguchi}},
  \bibinfo {author} {\bibfnamefont {J.}~\bibnamefont {Shan}},\ and\ \bibinfo
  {author} {\bibfnamefont {K.~F.}\ \bibnamefont {Mak}},\ }\bibfield  {title}
  {\bibinfo {title} {{Evidence of the fractional quantum spin Hall effect in
  moir{\'e} MoTe2}},\ }\href {https://doi.org/10.1038/s41586-024-07214-5}
  {\bibfield  {journal} {\bibinfo  {journal} {Nature}\ }\textbf {\bibinfo
  {volume} {628}},\ \bibinfo {pages} {522} (\bibinfo {year}
  {2024})}\BibitemShut {NoStop}%
\bibitem [{\citenamefont {Ghiasi}\ \emph {et~al.}(2025)\citenamefont {Ghiasi},
  \citenamefont {Petrosyan}, \citenamefont {Ingla-Ayn{\'e}s}, \citenamefont
  {Bras}, \citenamefont {Watanabe}, \citenamefont {Taniguchi}, \citenamefont
  {Ma{\~n}as-Valero}, \citenamefont {Coronado}, \citenamefont {Zollner},
  \citenamefont {Fabian}, \citenamefont {Kim},\ and\ \citenamefont {van~der
  Zant}}]{Ghiasi2025}%
  \BibitemOpen
  \bibfield  {author} {\bibinfo {author} {\bibfnamefont {T.~S.}\ \bibnamefont
  {Ghiasi}}, \bibinfo {author} {\bibfnamefont {D.}~\bibnamefont {Petrosyan}},
  \bibinfo {author} {\bibfnamefont {J.}~\bibnamefont {Ingla-Ayn{\'e}s}},
  \bibinfo {author} {\bibfnamefont {T.}~\bibnamefont {Bras}}, \bibinfo {author}
  {\bibfnamefont {K.}~\bibnamefont {Watanabe}}, \bibinfo {author}
  {\bibfnamefont {T.}~\bibnamefont {Taniguchi}}, \bibinfo {author}
  {\bibfnamefont {S.}~\bibnamefont {Ma{\~n}as-Valero}}, \bibinfo {author}
  {\bibfnamefont {E.}~\bibnamefont {Coronado}}, \bibinfo {author}
  {\bibfnamefont {K.}~\bibnamefont {Zollner}}, \bibinfo {author} {\bibfnamefont
  {J.}~\bibnamefont {Fabian}}, \bibinfo {author} {\bibfnamefont
  {P.}~\bibnamefont {Kim}},\ and\ \bibinfo {author} {\bibfnamefont {H.~S.~J.}\
  \bibnamefont {van~der Zant}},\ }\bibfield  {title} {\bibinfo {title} {Quantum
  spin hall effect in magnetic graphene},\ }\href
  {https://doi.org/10.1038/s41467-025-60377-1} {\bibfield  {journal} {\bibinfo
  {journal} {Nature Communications}\ }\textbf {\bibinfo {volume} {16}},\
  \bibinfo {pages} {5336} (\bibinfo {year} {2025})}\BibitemShut {NoStop}%
\bibitem [{\citenamefont {Maciejko}\ \emph {et~al.}(2009)\citenamefont
  {Maciejko}, \citenamefont {Liu}, \citenamefont {Oreg}, \citenamefont {Qi},
  \citenamefont {Wu},\ and\ \citenamefont {Zhang}}]{Maciejko09}%
  \BibitemOpen
  \bibfield  {author} {\bibinfo {author} {\bibfnamefont {J.}~\bibnamefont
  {Maciejko}}, \bibinfo {author} {\bibfnamefont {C.}~\bibnamefont {Liu}},
  \bibinfo {author} {\bibfnamefont {Y.}~\bibnamefont {Oreg}}, \bibinfo {author}
  {\bibfnamefont {X.-L.}\ \bibnamefont {Qi}}, \bibinfo {author} {\bibfnamefont
  {C.}~\bibnamefont {Wu}},\ and\ \bibinfo {author} {\bibfnamefont {S.-C.}\
  \bibnamefont {Zhang}},\ }\bibfield  {title} {\bibinfo {title} {Kondo effect
  in the helical edge liquid of the quantum spin hall state},\ }\href
  {https://doi.org/10.1103/PhysRevLett.102.256803} {\bibfield  {journal}
  {\bibinfo  {journal} {Phys. Rev. Lett.}\ }\textbf {\bibinfo {volume} {102}},\
  \bibinfo {pages} {256803} (\bibinfo {year} {2009})}\BibitemShut {NoStop}%
\bibitem [{\citenamefont {Schmidt}\ \emph {et~al.}(2012)\citenamefont
  {Schmidt}, \citenamefont {Rachel}, \citenamefont {von Oppen},\ and\
  \citenamefont {Glazman}}]{Schmidt12}%
  \BibitemOpen
  \bibfield  {author} {\bibinfo {author} {\bibfnamefont {T.~L.}\ \bibnamefont
  {Schmidt}}, \bibinfo {author} {\bibfnamefont {S.}~\bibnamefont {Rachel}},
  \bibinfo {author} {\bibfnamefont {F.}~\bibnamefont {von Oppen}},\ and\
  \bibinfo {author} {\bibfnamefont {L.~I.}\ \bibnamefont {Glazman}},\
  }\bibfield  {title} {\bibinfo {title} {Inelastic electron backscattering in a
  generic helical edge channel},\ }\href
  {https://doi.org/10.1103/PhysRevLett.108.156402} {\bibfield  {journal}
  {\bibinfo  {journal} {Phys. Rev. Lett.}\ }\textbf {\bibinfo {volume} {108}},\
  \bibinfo {pages} {156402} (\bibinfo {year} {2012})}\BibitemShut {NoStop}%
\bibitem [{\citenamefont {V\"ayrynen}\ \emph {et~al.}(2013)\citenamefont
  {V\"ayrynen}, \citenamefont {Goldstein},\ and\ \citenamefont
  {Glazman}}]{Vayrynen13}%
  \BibitemOpen
  \bibfield  {author} {\bibinfo {author} {\bibfnamefont {J.~I.}\ \bibnamefont
  {V\"ayrynen}}, \bibinfo {author} {\bibfnamefont {M.}~\bibnamefont
  {Goldstein}},\ and\ \bibinfo {author} {\bibfnamefont {L.~I.}\ \bibnamefont
  {Glazman}},\ }\bibfield  {title} {\bibinfo {title} {{Helical Edge Resistance
  Introduced by Charge Puddles}},\ }\href
  {https://doi.org/10.1103/PhysRevLett.110.216402} {\bibfield  {journal}
  {\bibinfo  {journal} {Phys. Rev. Lett.}\ }\textbf {\bibinfo {volume} {110}},\
  \bibinfo {pages} {216402} (\bibinfo {year} {2013})}\BibitemShut {NoStop}%
\bibitem [{\citenamefont {Buttiker}(1988)}]{Buttiker1988}%
  \BibitemOpen
  \bibfield  {author} {\bibinfo {author} {\bibfnamefont {M.}~\bibnamefont
  {Buttiker}},\ }\bibfield  {title} {\bibinfo {title} {Absence of
  backscattering in the quantum {Hall} effect in multiprobe conductors},\
  }\href {https://doi.org/10.1103/PhysRevB.38.9375} {\bibfield  {journal}
  {\bibinfo  {journal} {Phys. Rev. B}\ }\textbf {\bibinfo {volume} {38}},\
  \bibinfo {pages} {9375} (\bibinfo {year} {1988})}\BibitemShut {NoStop}%
\bibitem [{\citenamefont {Wen}(1994)}]{Wen1994}%
  \BibitemOpen
  \bibfield  {author} {\bibinfo {author} {\bibfnamefont {X.-G.}\ \bibnamefont
  {Wen}},\ }\bibfield  {title} {\bibinfo {title} {Impurity effects on chiral
  one-dimensional electron systems},\ }\href
  {https://doi.org/10.1103/PhysRevB.50.5420} {\bibfield  {journal} {\bibinfo
  {journal} {Phys. Rev. B}\ }\textbf {\bibinfo {volume} {50}},\ \bibinfo
  {pages} {5420} (\bibinfo {year} {1994})}\BibitemShut {NoStop}%
\bibitem [{\citenamefont {Kane}\ and\ \citenamefont {Fisher}(1995)}]{Kane1995}%
  \BibitemOpen
  \bibfield  {author} {\bibinfo {author} {\bibfnamefont {C.~L.}\ \bibnamefont
  {Kane}}\ and\ \bibinfo {author} {\bibfnamefont {M.~P.~A.}\ \bibnamefont
  {Fisher}},\ }\bibfield  {title} {\bibinfo {title} {Contacts and edge-state
  equilibration in the fractional quantum {Hall} effect},\ }\href
  {https://doi.org/10.1103/PhysRevB.52.17393} {\bibfield  {journal} {\bibinfo
  {journal} {Phys. Rev. B}\ }\textbf {\bibinfo {volume} {52}},\ \bibinfo
  {pages} {17393} (\bibinfo {year} {1995})}\BibitemShut {NoStop}%
\bibitem [{\citenamefont {Domaretskiy}\ \emph {et~al.}(2025)\citenamefont
  {Domaretskiy}, \citenamefont {Wu}, \citenamefont {Nguyen}, \citenamefont
  {Hayward}, \citenamefont {Babich}, \citenamefont {Li}, \citenamefont
  {Nguyen}, \citenamefont {Barrier}, \citenamefont {Indykiewicz}, \citenamefont
  {Wang}, \citenamefont {Gorbachev}, \citenamefont {Xin}, \citenamefont
  {Watanabe}, \citenamefont {Taniguchi}, \citenamefont {Hague}, \citenamefont
  {Fal\'{k}o}, \citenamefont {Grigorieva}, \citenamefont {Ponomarenko},
  \citenamefont {Berdyugin},\ and\ \citenamefont {Geim}}]{Domaretskiy25}%
  \BibitemOpen
  \bibfield  {author} {\bibinfo {author} {\bibfnamefont {D.}~\bibnamefont
  {Domaretskiy}}, \bibinfo {author} {\bibfnamefont {Z.}~\bibnamefont {Wu}},
  \bibinfo {author} {\bibfnamefont {V.~H.}\ \bibnamefont {Nguyen}}, \bibinfo
  {author} {\bibfnamefont {N.}~\bibnamefont {Hayward}}, \bibinfo {author}
  {\bibfnamefont {I.}~\bibnamefont {Babich}}, \bibinfo {author} {\bibfnamefont
  {X.}~\bibnamefont {Li}}, \bibinfo {author} {\bibfnamefont {E.}~\bibnamefont
  {Nguyen}}, \bibinfo {author} {\bibfnamefont {J.}~\bibnamefont {Barrier}},
  \bibinfo {author} {\bibfnamefont {K.}~\bibnamefont {Indykiewicz}}, \bibinfo
  {author} {\bibfnamefont {W.}~\bibnamefont {Wang}}, \bibinfo {author}
  {\bibfnamefont {R.~V.}\ \bibnamefont {Gorbachev}}, \bibinfo {author}
  {\bibfnamefont {N.}~\bibnamefont {Xin}}, \bibinfo {author} {\bibfnamefont
  {K.}~\bibnamefont {Watanabe}}, \bibinfo {author} {\bibfnamefont
  {T.}~\bibnamefont {Taniguchi}}, \bibinfo {author} {\bibfnamefont
  {L.}~\bibnamefont {Hague}}, \bibinfo {author} {\bibfnamefont {V.~I.}\
  \bibnamefont {Fal\'{k}o}}, \bibinfo {author} {\bibfnamefont {I.~V.}\
  \bibnamefont {Grigorieva}}, \bibinfo {author} {\bibfnamefont {L.~A.}\
  \bibnamefont {Ponomarenko}}, \bibinfo {author} {\bibfnamefont {A.~I.}\
  \bibnamefont {Berdyugin}},\ and\ \bibinfo {author} {\bibfnamefont {A.~K.}\
  \bibnamefont {Geim}},\ }\bibfield  {title} {\bibinfo {title} {Proximity
  screening greatly enhances electronic quality of graphene},\ }\href
  {https://doi.org/10.1038/s41586-025-09386-0} {\bibfield  {journal} {\bibinfo
  {journal} {Nature}\ }\textbf {\bibinfo {volume} {644}},\ \bibinfo {pages}
  {646} (\bibinfo {year} {2025})}\BibitemShut {NoStop}%
\bibitem [{\citenamefont {Wang}\ \emph {et~al.}(2013)\citenamefont {Wang},
  \citenamefont {Meric}, \citenamefont {Huang}, \citenamefont {Gao},
  \citenamefont {Gao}, \citenamefont {Tran}, \citenamefont {Taniguchi},
  \citenamefont {Watanabe}, \citenamefont {Campos}, \citenamefont {Muller},
  \citenamefont {Guo}, \citenamefont {Kim}, \citenamefont {Hone}, \citenamefont
  {Shepard},\ and\ \citenamefont {Dean}}]{Wang13}%
  \BibitemOpen
  \bibfield  {author} {\bibinfo {author} {\bibfnamefont {L.}~\bibnamefont
  {Wang}}, \bibinfo {author} {\bibfnamefont {I.}~\bibnamefont {Meric}},
  \bibinfo {author} {\bibfnamefont {P.~Y.}\ \bibnamefont {Huang}}, \bibinfo
  {author} {\bibfnamefont {Q.}~\bibnamefont {Gao}}, \bibinfo {author}
  {\bibfnamefont {Y.}~\bibnamefont {Gao}}, \bibinfo {author} {\bibfnamefont
  {H.}~\bibnamefont {Tran}}, \bibinfo {author} {\bibfnamefont {T.}~\bibnamefont
  {Taniguchi}}, \bibinfo {author} {\bibfnamefont {K.}~\bibnamefont {Watanabe}},
  \bibinfo {author} {\bibfnamefont {L.~M.}\ \bibnamefont {Campos}}, \bibinfo
  {author} {\bibfnamefont {D.~A.}\ \bibnamefont {Muller}}, \bibinfo {author}
  {\bibfnamefont {J.}~\bibnamefont {Guo}}, \bibinfo {author} {\bibfnamefont
  {P.}~\bibnamefont {Kim}}, \bibinfo {author} {\bibfnamefont {J.}~\bibnamefont
  {Hone}}, \bibinfo {author} {\bibfnamefont {K.~L.}\ \bibnamefont {Shepard}},\
  and\ \bibinfo {author} {\bibfnamefont {C.~R.}\ \bibnamefont {Dean}},\
  }\bibfield  {title} {\bibinfo {title} {{One-Dimensional Electrical Contact to
  a Two-Dimensional Material}},\ }\href
  {https://doi.org/10.1126/science.1244358} {\bibfield  {journal} {\bibinfo
  {journal} {Science}\ }\textbf {\bibinfo {volume} {342}},\ \bibinfo {pages}
  {614} (\bibinfo {year} {2013})}\BibitemShut {NoStop}%
\bibitem [{\citenamefont {Abanin}\ \emph {et~al.}(2007)\citenamefont {Abanin},
  \citenamefont {Novoselov}, \citenamefont {Zeitler}, \citenamefont {Lee},
  \citenamefont {Geim},\ and\ \citenamefont {Levitov}}]{Abanin07b}%
  \BibitemOpen
  \bibfield  {author} {\bibinfo {author} {\bibfnamefont {D.~A.}\ \bibnamefont
  {Abanin}}, \bibinfo {author} {\bibfnamefont {K.~S.}\ \bibnamefont
  {Novoselov}}, \bibinfo {author} {\bibfnamefont {U.}~\bibnamefont {Zeitler}},
  \bibinfo {author} {\bibfnamefont {P.~A.}\ \bibnamefont {Lee}}, \bibinfo
  {author} {\bibfnamefont {A.~K.}\ \bibnamefont {Geim}},\ and\ \bibinfo
  {author} {\bibfnamefont {L.~S.}\ \bibnamefont {Levitov}},\ }\bibfield
  {title} {\bibinfo {title} {{Dissipative Quantum Hall Effect in Graphene near
  the Dirac Point}},\ }\href {https://doi.org/10.1103/PhysRevLett.98.196806}
  {\bibfield  {journal} {\bibinfo  {journal} {Phys. Rev. Lett.}\ }\textbf
  {\bibinfo {volume} {98}},\ \bibinfo {pages} {196806} (\bibinfo {year}
  {2007})}\BibitemShut {NoStop}%
\bibitem [{\citenamefont {B\"uttiker}(1986)}]{Buttiker1986}%
  \BibitemOpen
  \bibfield  {author} {\bibinfo {author} {\bibfnamefont {M.}~\bibnamefont
  {B\"uttiker}},\ }\bibfield  {title} {\bibinfo {title} {Four-terminal
  phase-coherent conductance},\ }\href
  {https://doi.org/10.1103/PhysRevLett.57.1761} {\bibfield  {journal} {\bibinfo
   {journal} {Phys. Rev. Lett.}\ }\textbf {\bibinfo {volume} {57}},\ \bibinfo
  {pages} {1761} (\bibinfo {year} {1986})}\BibitemShut {NoStop}%
\bibitem [{\citenamefont {Groth}\ \emph {et~al.}(2014)\citenamefont {Groth},
  \citenamefont {Wimmer}, \citenamefont {Akhmerov},\ and\ \citenamefont
  {Waintal}}]{Groth2014}%
  \BibitemOpen
  \bibfield  {author} {\bibinfo {author} {\bibfnamefont {C.~W.}\ \bibnamefont
  {Groth}}, \bibinfo {author} {\bibfnamefont {M.}~\bibnamefont {Wimmer}},
  \bibinfo {author} {\bibfnamefont {A.~R.}\ \bibnamefont {Akhmerov}},\ and\
  \bibinfo {author} {\bibfnamefont {X.}~\bibnamefont {Waintal}},\ }\bibfield
  {title} {\bibinfo {title} {Kwant: a software package for quantum transport},\
  }\href {https://doi.org/10.1088/1367-2630/16/6/063065} {\bibfield  {journal}
  {\bibinfo  {journal} {New J. Phys.}\ }\textbf {\bibinfo {volume} {16}},\
  \bibinfo {pages} {063065} (\bibinfo {year} {2014})}\BibitemShut {NoStop}%
\bibitem [{\citenamefont {Han}\ \emph {et~al.}(2021)\citenamefont {Han},
  \citenamefont {Yang}, \citenamefont {Zhang}, \citenamefont {Wang},
  \citenamefont {Watanabe}, \citenamefont {Taniguchi}, \citenamefont {McEuen},\
  and\ \citenamefont {Ju}}]{Han2021}%
  \BibitemOpen
  \bibfield  {author} {\bibinfo {author} {\bibfnamefont {T.}~\bibnamefont
  {Han}}, \bibinfo {author} {\bibfnamefont {J.}~\bibnamefont {Yang}}, \bibinfo
  {author} {\bibfnamefont {Q.}~\bibnamefont {Zhang}}, \bibinfo {author}
  {\bibfnamefont {L.}~\bibnamefont {Wang}}, \bibinfo {author} {\bibfnamefont
  {K.}~\bibnamefont {Watanabe}}, \bibinfo {author} {\bibfnamefont
  {T.}~\bibnamefont {Taniguchi}}, \bibinfo {author} {\bibfnamefont {P.~L.}\
  \bibnamefont {McEuen}},\ and\ \bibinfo {author} {\bibfnamefont
  {L.}~\bibnamefont {Ju}},\ }\bibfield  {title} {\bibinfo {title} {Accurate
  measurement of the gap of
  $\mathrm{Graphene}/h\text{\ensuremath{-}}\mathrm{BN}$ moir\'e superlattice
  through photocurrent spectroscopy},\ }\href
  {https://doi.org/10.1103/PhysRevLett.126.146402} {\bibfield  {journal}
  {\bibinfo  {journal} {Phys. Rev. Lett.}\ }\textbf {\bibinfo {volume} {126}},\
  \bibinfo {pages} {146402} (\bibinfo {year} {2021})}\BibitemShut {NoStop}%
\bibitem [{\citenamefont {Ribeiro-Palau}\ \emph {et~al.}(2018)\citenamefont
  {Ribeiro-Palau}, \citenamefont {Zhang}, \citenamefont {Watanabe},
  \citenamefont {Taniguchi}, \citenamefont {Hone},\ and\ \citenamefont
  {Dean}}]{Palau2018}%
  \BibitemOpen
  \bibfield  {author} {\bibinfo {author} {\bibfnamefont {R.}~\bibnamefont
  {Ribeiro-Palau}}, \bibinfo {author} {\bibfnamefont {C.}~\bibnamefont
  {Zhang}}, \bibinfo {author} {\bibfnamefont {K.}~\bibnamefont {Watanabe}},
  \bibinfo {author} {\bibfnamefont {T.}~\bibnamefont {Taniguchi}}, \bibinfo
  {author} {\bibfnamefont {J.}~\bibnamefont {Hone}},\ and\ \bibinfo {author}
  {\bibfnamefont {C.~R.}\ \bibnamefont {Dean}},\ }\bibfield  {title} {\bibinfo
  {title} {Twistable electronics with dynamically rotatable heterostructures},\
  }\href {https://doi.org/10.1126/science.aat6981} {\bibfield  {journal}
  {\bibinfo  {journal} {Science}\ }\textbf {\bibinfo {volume} {361}},\ \bibinfo
  {pages} {690} (\bibinfo {year} {2018})}\BibitemShut {NoStop}%
\bibitem [{\citenamefont {Davies}\ \emph {et~al.}(1995)\citenamefont {Davies},
  \citenamefont {Larkin},\ and\ \citenamefont {Sukhorukov}}]{Davies1995}%
  \BibitemOpen
  \bibfield  {author} {\bibinfo {author} {\bibfnamefont {J.~H.}\ \bibnamefont
  {Davies}}, \bibinfo {author} {\bibfnamefont {I.~A.}\ \bibnamefont {Larkin}},\
  and\ \bibinfo {author} {\bibfnamefont {E.~V.}\ \bibnamefont {Sukhorukov}},\
  }\bibfield  {title} {\bibinfo {title} {Modeling the patterned two-dimensional
  electron gas: Electrostatics},\ }\href {https://doi.org/10.1063/1.359446}
  {\bibfield  {journal} {\bibinfo  {journal} {Journal of Applied Physics}\
  }\textbf {\bibinfo {volume} {77}},\ \bibinfo {pages} {4504} (\bibinfo {year}
  {1995})}\BibitemShut {NoStop}%
\end{thebibliography}%

\newpage

\clearpage

\onecolumngrid

\setcounter{figure}{0}

\renewcommand{\thefigure}{S\arabic{figure}}

\section*{Supplementary Information}

\subsection{Sample parameters}

\begin{figure}[hb!]
\centering
\includegraphics[width=1\columnwidth]{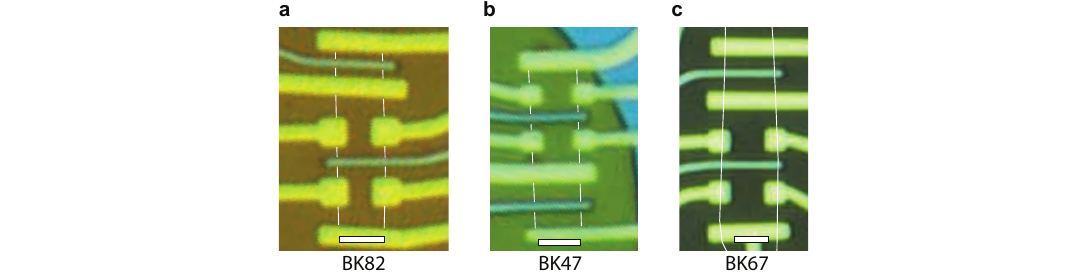}
\caption{
\textbf{Devices.}
\textbf{a-c}. Optical images of the Hall bar and 2-terminal devices on three different samples, with Au/Cr contacts in yellow and Pd top gate electrodes in grey.The white lines mark the pristine edges of the encapsulated graphene flakes. The scale bar in each image is 2 $\upmu$m. }
\label{figs1}
\end{figure}

Table \ref{Tab1} provides the thicknesses of the van der Waals assemblies for the four samples presented in this work. Fig. \ref{figs1} and Fig. \ref{fig4}c inset show optical images of the samples. Typical Hall mobilities range between 80 000 to 125 000$cm^2/V/s$ for sample BK82 and between 80 000 to 90 000$cm^2/V/s$ for sample BK67.  These values correspond to mean free paths of the order of the sample widths, indicating that mobilities are underestimated and that electronic transport is ballistic. 

\begin{table}[ht]
  \centering
  \caption{Layer thicknesses of van der Waals assemblies (Bi$_2$Se$_3$ / hBN$_\text{bottom}$ / hBN$_\text{top}$) for four samples.}
  \label{tab:vdw-thickness-bordered}
  \setlength{\arrayrulewidth}{0.6pt} 
  \renewcommand{\arraystretch}{1.2}  
  \begin{tabular}{|c|c|c|c|}
    \hline
    \textbf{  Sample  } & \textbf{Bi$_2$Se$_3$ (nm)} & \textbf{bottom hBN (nm)} & \textbf{top hBN (nm)}  \\
    \hline \hline
    BK41 & 64 & 3.7 & 29 \\
    \hline
    BK47 & 34.5 & 6.0 & 17 \\
    \hline
    BK67 & 64 & 5.2 & 28.3 \\
    \hline
    BK82 & 57.5 & 6.5 & 31 \\
    \hline
  \end{tabular}
  \label{Tab1}
\end{table}

\subsection{Reproducible helical edge transport in a second Hall bar device}  
\begin{figure*}[t!]
\includegraphics[width=1\linewidth]{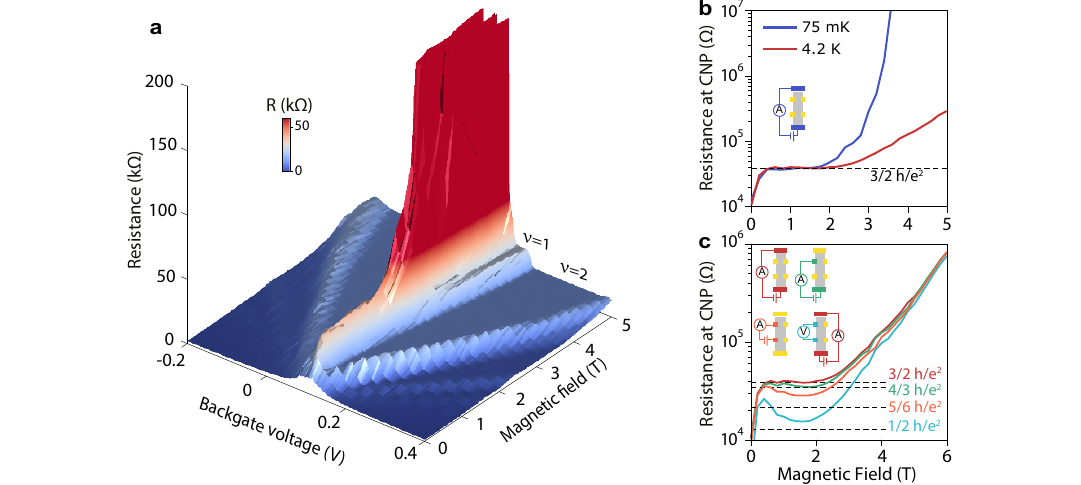}
\caption{  \textbf{Reproducible quantum Hall topological insulator with screened graphene.} \textbf{a.} Two-terminal resistance $R_{2t}$ as a function of magnetic field and back-gate voltage at $T = 75$~mK for sample BK47 with a bottom hBN spacer of 6 nm, showing quantized resistance plateau at charge neutrality associated with helical edge transport.  \textbf{b.} Temperature dependence of $R_{2t}$ at charge neutrality showing robust quantization down to 75~mK in the plateau region. \textbf{c.}  Two-terminal resistance curves at charge neutrality as a function of magnetic field at 4K for the configurations in the inset.}
\label{figS2}
\end{figure*}

In this section, we present data from an additional Hall bar device, confirming the reproducibility of the helical edge transport in our architecture. Fig. \ref{figS2} shows measurements from a Hall bar device on sample BK47 (see Fig. \ref{figs1}) similar to Hall bar device on BK87 discussed in the main text, except with an aspect ratio corresponding to roughly two-thirds of BK82. The magnetic field independent resistance plateau observed at charge neutrality in Fig.~\ref{figS2}a directly confirms the presence of helical edge transport in this sample. The resistance value at this plateau is quantized at $ {3}/{2}\times h/e^2 $, as evidenced by the linecuts of the two-terminal resistance at charge neutrality as a function of magnetic field and at two different temperatures in Fig. \ref{figS2}b. The resistance in the plateau region is virtually temperature independent, consistent with the expected metallic behavior, and shows a strong insulating trend at higher magnetic fields, consistent with a transition to a valley-polarized insulating phase. Despite the difference in sample size, the conductance is quantized at the expected value, consistent with helical edge transport. Changing the contact configuration yields distinct resistance plateaus (see Fig. \ref{figS2}c) that are in excellent agreement with the expected quantized values (black dashed lines) for two configurations (red and green), and slightly higher (by 15--30\%) for the orange and blue curves (four-terminal measurement of the orange section). 
This deviation can be attributed to the poor equilibration of the hole-like mode with the contact, as already discussed in the respective Methods section. While the simplistic model presented therein does not allow for resistance values outside of the two limiting values ($t = 0$ and $t = 1$), this emerges as an artifact of the approximation of constant $t$ values for all contacts. A more realistic model that takes into account the fluctuations of the hole transmission of each individual contact does explain the observed resistance values. For example, the value of resistance corresponding to the cyan curve in Fig.~\ref{figS2}c can assume any value in the range $[3R_Q / 4, R_Q]$ depending on the transmissions of individual contacts. In general, it is the collective information about multiple resistance probes on the same sample that reveals the actual hole transmission values for each contact.  The corresponding detailed analysis will be published elsewhere.    

\subsection{Helical-chiral junctions in a 2-terminal device}

\begin{figure*}[t!]
\includegraphics[width=1\linewidth]{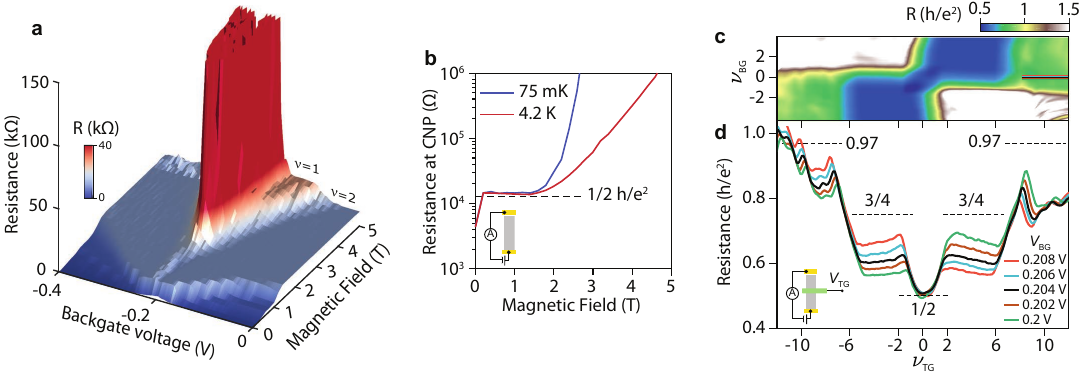}
\caption{  \textbf{Helical edge transport and selective backscattering in a 2-terminal device.} \textbf{a.} Two-terminal resistance $R_{2t}$ as a function of magnetic field and back-gate voltage at $T = 75$~mK, measured on the two terminal device on sample BK82. \textbf{b.} Temperature dependence of $R_{2t}$ at charge neutrality showing robust quantization down to 75~mK in the plateau region. \textbf{c.} Colormap of the two-terminal resistance measured in the configuration shown in the inset at 1.6~T and 4K, as a function of the top-gated region filling factor \(\nu_{\mathrm{TG}}\) and the back gate-only region filling factor \(\nu_{\mathrm{BG}}\). \textbf{d.}  Linecuts at fixed back-gate filling factors, $\nu_{\mathrm{BG}} = 0$ (black) and nearby values. Dashed lines indicate expected values.}
\label{figS3}
\end{figure*}

In this section we present evidence of helical edge transport through a chiral--helical junction in a 2-terminal device on sample BK82 in Fig. \ref{figs1}. In Fig. \ref{figS3}a, we first demonstrate uninterrupted helical edge transport in this device, while the top gate is set to isodensity. Similar to the Hall bar device, the charge neutral point at lower fields is conducting. The linecuts in resistance at charge neutrality (see Fig. \ref{figS3}b) as a function of magnetic field reveal virtually temperature-independent plateuas at low magnetic fields suggesting helical edge transport. The resistance value on the plateau is very close to the expected value of of 1/2 $h/e^2$ (off by $\sim1$k$\Omega$ or 8\%).

 Evidence of selective helical edge transport in this 2-terminal device is presented in Fig.~\ref{figS3}c,d. In Fig. \ref{figS3}c, we plot a two-dimensional resistance map as a function of the top- and back-gate filling factors, $\nu_{\mathrm{TG}}$ and $\nu_{\mathrm{BG}}$, respectively, for the configuration shown. The corresponding linecut at fixed $\nu_{\mathrm{BG}} = 0$ is displayed in Fig. \ref{figS3}d  (black curve), together with traces near $\nu_{\mathrm{BG}} = 0$. All curves exhibit a global minimum at $\nu_{\mathrm{TG} = 0}$, with the $\nu_{\mathrm{BG}} = 0$ curve showing the largest resistance value at the minimum. This resistance minimum corresponds to the expected value of $1/2\,h/e^2$ for helical edge transport. Away from $\nu_{\mathrm{TG}} = 0$, all curves display a resistance jump followed by a plateau at $\nu_{\mathrm{TG}} = \pm 2$. This jump in resistance is associated with the backscattering of one helical edge channel at the chiral-helical junction. Interestingly, unlike the $\nu_{\mathrm{BG}} = 0$ curve (black), these plateaus in other curves are not symmetric between the electron and hole sides. 

To understand this asymmetry, we consider bulk leakage that arises away from charge neutrality due to bulk states forming near the bulk-edge boundary, associated with broadened spin-split Landau levels at $\nu_{\mathrm{BG}} = \pm 1$ \cite{Abanin07b}. These bulk states provide a parallel spin-polarized channel (electron-like or hole-like) in addition to the helical edge channels. For $\nu_{\mathrm{BG}} > 0$ (orange and blue curves) and $\nu_{\mathrm{TG}} > 0$, the additional bulk electron channel is transmitted together with the electron-like helical edge channel for $\nu_{\mathrm{TG}} > 0$, thus lowering the resistance. In contrast, for $\nu_{\mathrm{TG}} < 0$, this channel is backscattered along with the electron-like helical edge channel, leading to an increase in resistance. For $\nu_{\mathrm{BG}} < 0$ (brown and green curves), the opposite behavior is observed, consistent with the described bulk leakage picture.

Following the methods presented in the main text for selective transmission and equilibration of helical edge states in the Hall bar device, we also calculated the expected resistance value for the 2-terminal device geometry when the bulk is tuned to the QHTI phase $\nu_{\mathrm{BG}}  = 0$ while the top gate region is tuned to $\nu_{\mathrm{TG}}  = \pm2$ and $\nu_{\mathrm{TG}}  = \pm6$ and obtained values of $3/4 h/e^2$ and $0.97 h/e^2$, respectively (dashed lines in Fig. \ref{figS3}d). 

\subsection{Insulating state due to Moir\'e induced gap}
Graphene aligned with hexagonal boron nitride (hBN) forms a moir\'e superlattice that modifies its band structure and opens an energy gap at the charge neutrality~\cite{Han2021,Palau2018}. The magnitude of this gap depends on the relative twist angle $\theta$ between the two layers, reaching its maximum value  at $\theta = 0^\circ$~\cite{Palau2018}. The gap remains substantial for $\theta \lesssim 3^\circ$, but decreases rapidly at larger twist angles, where it has a negligible effect on the band structure. The presence of such a moir\'e-induced gap at charge neutrality suppresses helical edge transport, irrespective of screening or even under extreme magnetic fields \cite{Young14}.

\begin{figure*}[h!]
\includegraphics[width=1\linewidth]{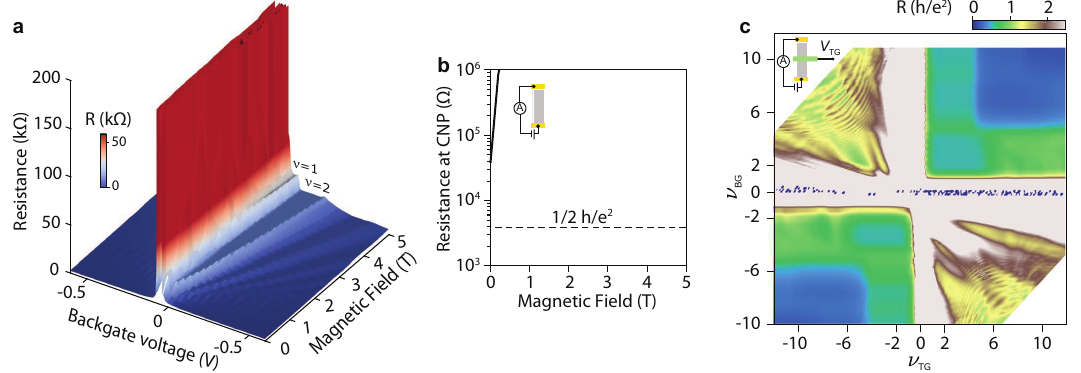}
\caption{  \textbf{Moir\'e gap conceals graphene quantum Hall topological insulator.} \textbf{a.} Two-terminal resistance $R_{2t}$ as a function of magnetic field and back-gate voltage at $T = 4$~K, in a 2-terminal device (BK67 in Fig. \ref{figs1}) with a Moire gap due to the hBN graphene interface.  \textbf{b.} Two-terminal resistance at charge neutrality at 4~K  showing a rapid transition to an insulating state as a function of magnetic field. \textbf{c.} Colormap of the two-terminal resistance measured in the configuration shown in the inset at 1.6~T, as a function of the top-gated region filling factor \(\nu_{\mathrm{TG}}\) and the back gate-only region filling factor \(\nu_{\mathrm{BG}}\).}
\label{figS4}
\end{figure*}

In our fabrication process, unintentional or partial self-alignment occurred as the graphene or bottom hBN flake shifted at the elevated temperatures used for assembly. As a result, a significant fraction of our samples displayed a moir\'e gap, evidenced by a highly insulating state at charge neutrality (see Fig. \ref{figS4}).

The formation of a moir\'e pattern can be avoided by deliberately misaligning the graphene and hBN lattices, which requires knowledge of their crystallographic orientations. Predicting these orientations, however, can be challenging due to the irregular shapes of exfoliated hBN flakes. To overcome this, we selected hBN flakes with long, straight edges, which typically follow the crystallographic axes. These edges were then intentionally misaligned with the graphene edge by approximately $15^\circ$. The resulting devices exhibited robust signatures of helical edge transport, with no evidence of moir\'e-related effects.

\subsection{Model for helical edge transport in graphene QHTI}

In our simulations, we use a honeycomb lattice defined by basis vectors $\mathbf{a}_1 = a \sqrt{3} \mathbf{e_x}$ and $\mathbf{a}_2 =  a (\sqrt{3} \mathbf{e_x} + 3\mathbf{ e_y})/2$, where $a = 1.42~\textup{\AA} $ is the lattice constant of graphene.  
The scattering region is a system of irregular shape with $N = 644622$ sites, based on a rectangle of size $L_x \times L_y = 1800 a \times 300a $ as discussed in the main text.
It is described by the real-space Hamiltonian 
\begin{eqnarray} \label{eq:ham}
H = \sum_{n= 1}^{N} \mu_n^{\rm tot} c_n^{\dagger} \sigma_0 c_n + \sum_{n= 1}^{N} V_z c_n^{\dagger} \sigma_z c_n
- \sum_{\braket{nm}} t_{nm} c_n^{\dagger} \sigma_0 c_m^{} + \rm{ h.c.},
\end{eqnarray}
that is written in the basis $c_n = (c_{n, \uparrow}, c_{n, \downarrow})$, where $c_{n, s}$ and $c_{n, s}^{\dagger}$ are the annihilation and creation, respectively operator of an electron at site $n$ and spin $s=\uparrow,\downarrow$.
The Pauli matrix $\sigma$ represents the spin degree of freedom, and $V_z = 0.1~\rm eV$ is the Zeeman splitting in the \textit{z}-direction of the lowest Landau Levels arising from Coulomb interactions. 
The nearest neighbor hopping amplitude $t_{nm}$ is assumed to be constant in the absence of the magnetic field, and equals $t = 2.8~\rm eV$ in graphene. 
Lastly, $\mu_n^{\rm tot} \equiv \mu^{\rm tot} (x_n,y_n)$ is the total chemical potential of site $n$ at position $(x_n,y_n)$.

As described in the Methods section of the main text, the total chemical potential of site $n$ at position $(x_n,y_n)$ is a combination of several terms
\begin{equation}\label{eq:chem_pot}
 \mu^{\rm tot} (x_n,y_n)  =  \mu_0 + \nu^b  (x_n,y_n)+ 
  \eta (x_n,y_n) \big(V_g+  \nu^c  (x_n,y_n) \big).
\end{equation}
Here, $\mu_0 = 0$ is the reference chemical potential, constant throughout the system, and $\nu^b  (x_n,y_n)$ represents the bulk disorder potential originating from imperfections in the substrate. 
This type of disorder is present in the entire system, and its value $\eta^b (x_n,y_n) $ at site $n$ is drawn independently from a uniform distribution $[-W^{b}/2, W^{b}/2]$, where $W^b = 0.5 ~\rm{eV}$$ \approx 0.18 t$. 

The remaining terms in Eq.~\eqref{eq:chem_pot} represent the contributions to the chemical potential from the contacts attached to longer edges.
For the six-terminal setup discussed in the Methods section of the main text, these contacts were used to measure voltages, i.e., they were voltage gates. 
The first term, $\eta(x_n, y_n) V_g$, represents the change in the chemical potential due to patterning these contacts; here $V_g = 0.2~\rm eV$ is the voltage gate potential.
It is taken to be smaller than the chemical potential of the contacts in order to capture imperfections in the coupling between contacts and the scattering region.  
The function $\eta(x_n, y_n) = \sum_{i} \eta_i(x_n, y_n)$ is defined as a sum over contacts $i$. Each $\eta_i(x_n, y_n)$ is defined such that $\eta_i(x_n, y_n) \rightarrow 0$ in the bulk of the sample, and $\eta_i(x_n, y_n) \rightarrow 1$ in the region where the $i$-th contact is patterned. 
Hence, $\eta_i(x_n, y_n)$ describes the spatial variation of the chemical potential due to contact $i$ with boundaries $r_{L, i} < x < r_{R, i}$ and $r_{B,i} <y < r_{T, i}$, and is given by~\cite{Davies1995} 
\begin{equation} \label{eq:chem_pot_var}
\eta_i (x_n,y_n) =g(x_n- r_{L, i} , y_n -r_{B, i}) + g(x_n-r_{L, i}, r_{T, i}-y_n) + 
g(r_{R,i}-x_n, y_n - r_{B,i})+g(r_{R, i}-x_n, r_{T,i}-y_n),
\end{equation}
with $g(u, v) = \frac{1}{2\pi} \arctan{\frac{uv}{dR}}$ and $R = \sqrt{u^2+v^2+d^2}$. 
The parameter $d$ represents the distance in the \textit{z}-direction between a patterned gate and the site at which we evaluate the chemical potential; here we choose $d = 2$.
To simplify the simulation, we assume that all four contacts attached to longer edges have the same width $w = r_{R, {i}}-r_{L, {i}}$.

The last term in Eq.~\eqref{eq:chem_pot} is $\eta(x_n, y_n) \nu^c (x_n, y_n)$ and captures imperfections in the sample induced by patterning the contacts.
Here, $\nu^c (x_n, y_n)$ is drawn independently for each site $n$ from a uniform distribution $[-W^{c}/2, W^{c}/2]$, with $W^c = 7~ \rm eV$$=2.5t$.
In practice, we add this contribution only for those sites for which $\nu^c (x_n, y_n) > 0.001$.
Notice that we work in the experimentally realistic regime when $W^{c}>W^{b}$, which allows propagating edge modes to equilibrate when these leads are sufficiently wide.

To model the effect of a magnetic field $\mathbf{B} = \nabla \times \mathbf{A}$ on the system, we use the vector potential in the Landau gauge $\mathbf{A} = (-By, 0,0)$.
Then, the hopping between nearest neighboring sites $n$ and $m$ is defined through the Peierls substitution rule as
\begin{equation}
t_{nm} = t \exp[ i \frac{2\pi (x_n-x_m) (y_n+y_m) B a^2}{2\phi_0} ],
\end{equation}
where $\phi_0 = h/e$ is the magnetic flux quantum and $B = 100 ~\rm T$ is the magnetic field strength.
From here, we set the magnetic flux per unit cell to be $\phi = B \times \frac{3\sqrt{3} a^2}{2} = 0.0012 \phi_0$. Its corresponding magnetic length, that characterizes the spatial extent of the Landau levels, is $l_B = \sqrt{\hbar/eB} \approx 18a$.

So far, we have explained all parameters of the scattering region. 
In the following, we describe how we simulate electrical contacts. 
To reduce reflection at interfaces due to mismatch in lattices, we model both current and voltage gates as doped, disorder-free graphene with a large number of propagating states at the Fermi level.
These gates can be described by Eq.~\eqref{eq:ham} with parameters $\mu_0^{\rm lead} = 0.9t=2.52\rm eV$, $V_g = 0$, $W^b =W^c= 0$, $V_z =0$ and $t_{nm} = t = 2.8~\rm eV$.
Each of these contacts is considered semi-infinite in the direction perpendicular to the interface of this contact and the system.
%

\subsection{Two-terminal resistance for the ``green'' setup in Fig.~\ref{figS2}\textbf{c}}

\begin{figure}[tb!]
\centering
    \includegraphics[width=1\linewidth]{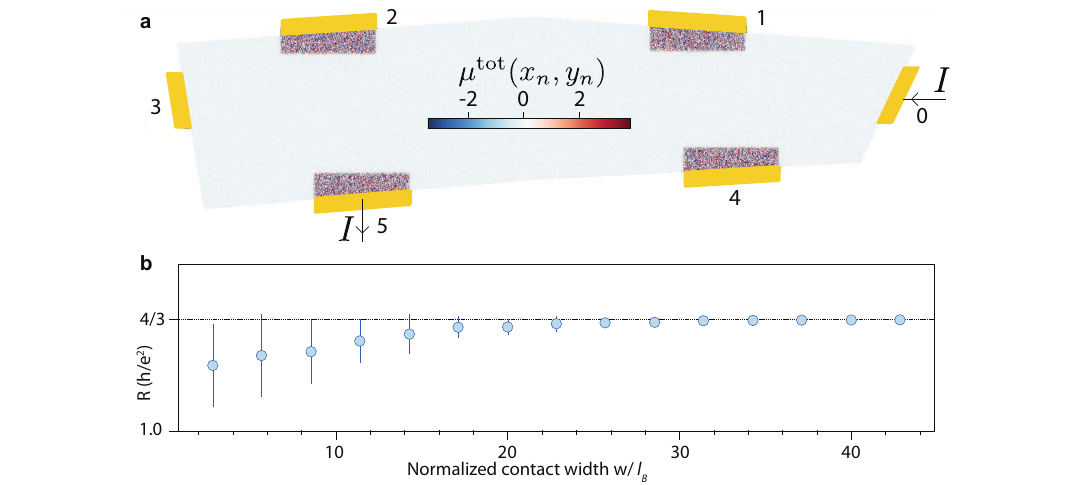}
\caption{Two-terminal resistance simulation of the ``green'' setup shown top-right inset of Fig.~\ref{figS2}{c}. \textbf{a.} The 6-terminal scattering setup where $i = 0.5$ denote two current leads, and $i = 1,2,3,4$ are voltage contacts. 
The color map represents the local potential given by Eq.~\eqref{eq:chem_pot}.
This setup simulates the schematics shown in the top-right inset of Fig.~\ref{figS2}{c}.
\textbf{b.} The resulting $R_{2t}$ as a function of the width $w$ of the contacts $ i =1,2,4, 5$. 
We use parameters $\mu_ 0 = 0$, $V_g = 0.2 \rm eV$, $W_A^c = 7 \rm eV$, $W_A^b = 0.5 \rm eV$, 
$t = 2.8\rm{eV}$, $B = 100 \rm T$, averaged over $N_{\rm dis} = 35$ realizations.}
\label{fig:conf1}
\end{figure}

In Fig.~\ref{fig:conf1}a, we illustrate the six-terminal scattering setup depicted schematically in the insets of Fig.~\ref{figS2}{b,c}. 
%
Here, the color bar indicates the total chemical potential of the scattering region for one disorder realization. 
In proximity to the side contacts $i = 0,3$ we choose $V_g, W^c = 0$ because in experiment, these contacts are patterned over the entire length of their corresponding side edges.
Hence, their size is much larger than the mean free path implying a perfect absorption of helical modes from the sample. 
In our simulation, we assume these contacts are $\approx L_y/2$ wide. 

In the configuration discussed in the main text, contacts $i=0$ and $i=3$ correspond to current probes. 
In the following, we discuss a different experimental configuration that corresponds to the green-contact setup, top right of the inset of Fig.~\ref{figS2}{c}, and orange setup in Fig.~\ref{fig2}{b} in the main text.
This configuration is depicted in Fig.~\ref{fig:conf1}{a}, where the current passes through contacts $i = 0,5$ and all other contacts are voltage gates.
Like for the transport geometry discussed in the main text, it was also experimentally observed that wider contacts are necessary for better quantization of the two-terminal resistance in this scattering setup.

To simulate the corresponding two-terminal resistance $R_{2t}$, we calculate the conductance matrix $G$ using Kwant~\cite{Groth2014}.
We focus on transport properties at $E_F = 0$. 
In order to determine the voltages at each contact, we solve the equation $\mathbf{V} = G^{-1} \mathbf{I}$, where $\mathbf{I}$ and $\mathbf{V}$ are current and voltage vectors with six entries corresponding to the six contacts.
Because currents between two contacts depend on their voltage differences 
we can fix one unknown by
assuming a zero voltage at the outgoing current terminal. 
The current at this terminal is deduced from Kirchhoff's law of current conservation, and equals minus the sum of all other currents.
Note that the current through the voltage contacts equals zero, by definition.

The resulting set of equations read $\tilde{\mathbf{V}} = \tilde{G}^{-1} \tilde{\mathbf{I}}$, where $\tilde{\mathbf{I}}$ and $\tilde{\mathbf{V}}$ have only five entries corresponding to one current and four voltage contacts.
After solving the matrix equation $\tilde{\mathbf{V}} = \tilde{G}^{-1} \tilde{\mathbf{I}}$, we can calculate the two-terminal resistance as 
\begin{equation}
R_{2t} = \frac{V_{i = \rm current \; contact}}{I},
\end{equation} 
where $I$ is the strength of the current passing through the system.
For both scattering setups simulated in this work and represented in Figs.~\ref{fig5} and~\ref{fig:conf1}, $i = 0$ corresponds to the current contact.

For the six-terminal configuration shown in Fig.~\ref{fig:conf1}{a}, there are 
two nonequivalent paths $0 \to 1 \rightarrow 2 \rightarrow 3 \rightarrow 5$ and $0 \rightarrow 4 \rightarrow 5$ that are connected in parallel, and along which helical modes propagate.
Assuming that the helical modes fully equilibrate in proximity to the contacts $i = 1,2,3,4$, the total resistance along the path $0 \rightarrow 1 \rightarrow 2 \rightarrow 3 \rightarrow 5$  is $4 R_0$, and $2R_0 $ along the path $0 \rightarrow 4 \rightarrow 5$. 
Thus, the total resistance of the sample at $E_F = 0$ is expected to be $R_{2t} = 4R_0/3$ in this configuration.
In Fig.~\ref{fig:conf1}{b}, we show how the two-terminal resistance reaches precisely that quantized value only upon increasing the width $w$ of contacts $i = 1,2,4,5$.
This is consistent with our findings described in the main text: wider leads allow full equilibration of the helical edge states.

\subsection{Two-terminal resistance addition}

\begin{figure*}[tb!]
\includegraphics[width=\textwidth]{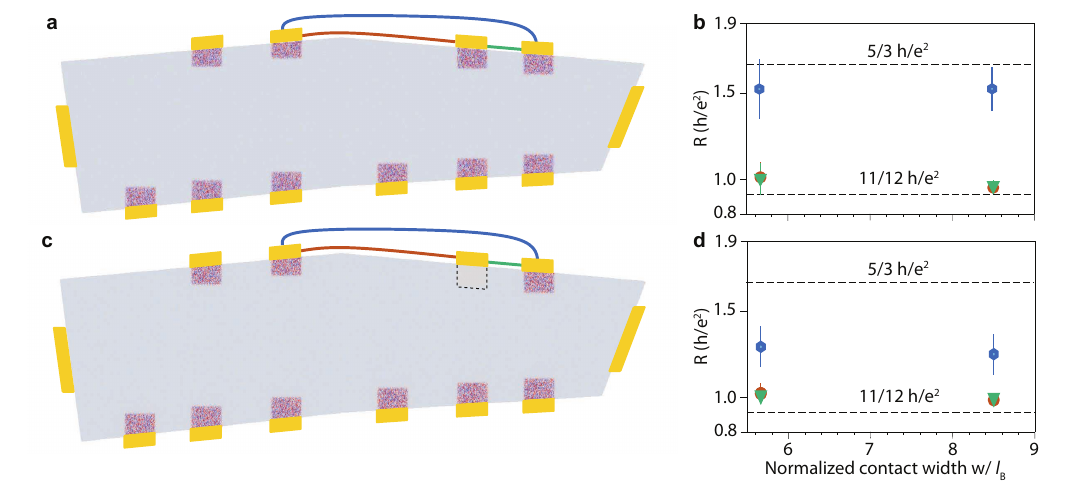}
\caption{
Effect of removing contact disorder on quantization.
\textbf{a.} 12-terminal scattering setup with blue, green, and orange lines indicating pairs of terminals used for measuring the two-terminal resistance, color coded as in Fig.~\ref{fig3}c. The color map is the same as in Fig.~\ref{fig:conf1} representing the disorder potential Eq.~\eqref{eq:chem_pot}. In this setup, we consider that all voltage contacts have a surrounding disordered region on top of a smooth gate potential depletion.
\textbf{b.} The corresponding disorder-averaged $R_{2t}$ as a function of $w$ for the fully disordered limit for the system in panel \textbf{a}.  
Larger $w/\ell_B$ bring the average $R_{2t}$ closer to the expected quantization values, indicated by dashed lines.
\textbf{c.} To analyze the effect of disorder we simulate the same 12-terminal scattering setup as in \textbf{a} with the difference that we turn off the disorder, but not the gating, from the intermediate top contact.
The gating is non-zero within the black dashed region, and it spatially separates the two counter propagating channel, but now contact-induced disorder, which helps equilibrate these two channels, is absent.
\textbf{d.} The resulting, disorder-averaged, $R_{2t}$ as a function of $w/\ell_B$ for the scattering shown in \textbf{d}.
We observe that the average $R_{2t}$ in \textbf{b} is closer to the expected quantized values compared to \textbf{d}.
This supports our conclusion that contact disorder is key to equilibrate the edge states and observe quantization.
We use parameters $\mu_ 0 = 0$, $V_g = 0.2 \rm eV$, $W_A^c = 7 \rm eV$, $W_A^b = 0.5 \rm eV$, $t = 2.8\rm{eV}$, $B = 100 \rm T$, and $N_{\rm dis} = 25$ disorder realizations.}
\label{fig:many_Leads}
\end{figure*}

In addition to a six-terminal transport geometry, in the main text we have presented experimental results for a transport geometry with as many as 12 terminals, see inset of Fig.~4\textbf{c}. 
Considering the short-circuited contacts as one (see black lines in the inset of Fig.~\ref{fig4}{c}), we illustrate one such transport geometry in Fig.~\ref{fig:many_Leads}{a}. If features six contacts attached to the top edge, four attached to the bottom edge and two at the left and right contacts.
We use this geometry to study the two-terminal resistance between two nearest neighboring contacts, indicated by  orange and green lines in Fig.~\ref{fig:many_Leads}{a}, and the two-terminal resistance between two terminals that are second nearest neighbors, indicated by the blue line in Fig.~\ref{fig:many_Leads}{a}.

Experimentally, we observe that the two-terminal resistance between neighboring contacts is slightly larger than the theoretical value $R_{2t}^{\rm th} = 11R_0/12$  obtained under the assumption that $T_{ij} \rightarrow 1$.
Concurrently, the two-terminal resistance between second nearest neighboring contacts is measured to be smaller than its theoretical value 
$R_{2t}^{\rm th} = 5R_0/3$.

To determine whether this behavior originates from disorder-induced scattering of the counter-propagating states at the electrical contacts, we simulate $R_{2t}$ as a function of contacts width $w$.
Due to the increased number of electrical contacts, we cannot reach contact widths $w$ as large as those simulating the six-terminal geometry under assumption that the scattering region remains the same size ($N = 644622$ sites).
Hence, in the following simulations we choose the mentioned $N$ and reduce the contact widths to $w / l_B = 5.67, 8.5$, i.e., values that are smaller than in the experiment.
Nevertheless, our simulations capture well the experimental observations.
Other parameters read $\mu_ 0 = 0$, $V_g = 0.2~\rm eV$, $W_A^c = 7~ \rm eV$, $W_A^b = 0.5~\rm eV$, $t = 2.8~\rm{eV}$, $B = 100~\rm T$, and $N_{\rm dis} = 25$.

In Fig.~\ref{fig:many_Leads}{b} we show how the two-terminal resistance changes as a function of $w$ for three different pairs of electrical contacts indicated by the blue, green, and orange lines in Fig.~\ref{fig:many_Leads}{a}.
For the given contacts widths, we see that the disorder-averaged $R_{2t}$ corresponding to pairs of nearest neighboring contacts (orange and green) is larger than theoretical value \ $R_{2t}^{\rm th} = 11R_0/12$. 
We observe that increasing $w$ brings these two-terminal resistances closer to the theoretical value, in agreement with wider contacts promoting $T_{ij} \rightarrow 1$.
Concomitantly, our simulations reveal that $T_{ij} < 1$ reduces the two-terminal resistance between second nearest neighboring contacts compared to the theoretical value  $R_{2t}^{\rm th} = 5R_0/3$.
Moreover, while $R_{2t}$ changes little for the contacts widths $w$ considered in the simulations, we see that wider terminals promote smaller deviations from the quantized value.

To study the effect disorder has on the two-terminal resistance, we eliminate the disorder in proximity of one of these contacts while keeping the gate potential, see top side of the sample in Fig.~\ref{fig:many_Leads}{c}. 
In Fig.~\ref{fig:many_Leads}{d}, we see that the two-terminal resistance for the blue configuration are farther away from the theoretical values compared to the fully disordered case shown in Fig.~\ref{fig:many_Leads}{b}.
This supports our conclusion that contact disorder is key to equilibrate the edge states and observe quantization.

\end{document}